\documentclass[aps,prd,10pt,nofootinbib,twocolumn,superscriptaddress,floatfix,notitlepage]{revtex4-2}

\usepackage{graphicx}
\usepackage{float}
\usepackage{placeins}

\usepackage{amsmath,amsfonts,amssymb}
\usepackage[dvipsnames]{xcolor}
\usepackage{verbatim}
\usepackage{enumitem}
\usepackage{aas_macros}
\usepackage[breaklinks,colorlinks,urlcolor=MidnightBlue,citecolor=WildStrawberry,linkcolor=Fuchsia]{hyperref}

\newcommand{\Msun}{M_\odot}
\newcommand{\td}{{\rm d}}
\newcommand{\pd}{{\partial}}

\newcommand{\sinc}{{\rm sinc}}

\newcommand{\be}{\begin{equation}}
\newcommand{\ee}{\end{equation}}
\newcommand{\bea}{\begin{equation} \begin{aligned}}
\newcommand{\eea}{\end{aligned} \end{equation}}

\def\lsim{\mathrel{\raise.3ex\hbox{$<$\kern-.75em\lower1ex\hbox{$\sim$}}}}
\def\gsim{\mathrel{\raise.3ex\hbox{$>$\kern-.75em\lower1ex\hbox{$\sim$}}}}

\begin{document}

\title{The Heavy Tailed Non-Gaussianity of the Supermassive Black Hole \\ Gravitational Wave Background}

\author{Juhan Raidal}
\email{juhan.raidal@kbfi.ee}
\affiliation{Laboratory of High Energy and Computational Physics, NICPB, R{\"a}vala 10, Tallinn, 10143, Estonia}
\affiliation{Department of Cybernetics, Tallinn University of Technology, Akadeemia tee 21, 12618 Tallinn, Estonia}

\author{Juan Urrutia}
\email{juan.urrutia@kbfi.ee}
\affiliation{Laboratory of High Energy and Computational Physics, NICPB, R{\"a}vala 10, Tallinn, 10143, Estonia}
\affiliation{Department of Cybernetics, Tallinn University of Technology, Akadeemia tee 21, 12618 Tallinn, Estonia}

\author{Ville Vaskonen}
\email{ville.vaskonen@kbfi.ee}
\affiliation{Laboratory of High Energy and Computational Physics, NICPB, R{\"a}vala 10, Tallinn, 10143, Estonia}
\affiliation{Dipartimento di Fisica e Astronomia, Universit\`a degli Studi di Padova, Via Marzolo 8, 35131 Padova, Italy}
\affiliation{Istituto Nazionale di Fisica Nucleare, Sezione di Padova, Via Marzolo 8, 35131 Padova, Italy}

\author{Hardi Veerm\"ae}
\email{hardi.veermae@cern.ch}
\affiliation{Laboratory of High Energy and Computational Physics, NICPB, R{\"a}vala 10, Tallinn, 10143, Estonia}

\begin{abstract}
We study the non-Gaussian features of the gravitational wave (GW) background generated by a population of inspiraling supermassive black hole (SMBH) binaries. We show that the SMBH GW amplitude distribution (GWAD) features a universal heavy power-law tail $\propto A^{-4}$, while the low-amplitude tail depends on the SMBH merger rate and the energy-loss mechanisms of the binaries. The distribution of the induced timing residuals inherits this heavy tail. As a result, the ensemble averaged statistical moments of order three and higher diverge, limiting their usefulness as measures of non-Gaussianity, and the GW background from SMBH binaries exhibits the single loud source principle, according to which the strongest signals are more likely to be caused by a small number of loud sources. We confirm that the variance-averaged Gaussian approximation accurately describes the timing residual statistics. This approximation justifies a factored likelihood structure that combines standard Gaussian-process PTA posteriors with the non-Gaussian population prior, enabling consistent incorporation of non-Gaussian effects into SMBH model inference. We provide a fast and flexible Python implementation to compute the distribution of timing residuals from a given SMBH merger rate or GWAD. 
\end{abstract}

\maketitle

\section{Introduction}
\label{sec:intro}

Multiple pulsar timing array (PTA) experiments have reported compelling evidence for a nHz stochastic gravitational wave (GW) background~\cite{NANOGrav:2023gor, EPTA:2023fyk, Reardon:2023gzh, Xu:2023wog}. The leading interpretation is that this signal arises from a population of supermassive black hole (SMBH) binaries that are created in galaxy mergers and radiate GWs as they inspiral~\cite{Rajagopal:1994zj, Wyithe:2002ep, Sesana:2004sp, NANOGrav:2023gor, NANOGrav:2023hfp, EPTA:2023xxk, Ellis:2023dgf}. Alternatively, the background can arise from various early-Universe processes~\cite{NANOGrav:2023hvm, EPTA:2023xxk, Ellis:2023oxs}.

A cosmological stochastic GW background is typically modeled as a Gaussian random process because it arises from the superposition of signals emitted by a large number of independent, causally disconnected regions. By the central limit theorem, the sum of many uncorrelated contributions tends toward Gaussian statistics, largely independent of the detailed properties of the individual sources. Motivated by this, PTA analyzes commonly describe the GW background through a Gaussian random process that is isotropic, unpolarized, and static (see e.g.~\cite{Taylor:2016ftv,Taylor:2021yjx}). This description applies at the level of the ensemble, while cosmic variance can induce apparent anisotropies in individual realizations even when the process is statistically isotropic~\cite{Domcke:2025esw}.

In contrast, the nHz GW background from SMBH binaries represents one particular realization stemming from a finite population of binaries. Although the total number of binaries that emit in the nHz band can be very large, the vast majority contribute negligibly to the GW background. Instead, it is likely that the background is dominated by a relatively small number of loud binaries, some of which may become individually resolvable as the PTA sensitivity improves~\cite{Sesana:2008xk,Rosado:2015epa,Kelley:2017vox,Mingarelli:2026kjw}. The background is static because these binaries are far from coalescence, and their emission is nearly monochromatic but exhibits significant anisotropies~\cite{Taylor:2020zpk,Gardiner:2023zzr,Sato-Polito:2023spo,Raidal:2024tui,Sah:2024oyg,Sah:2025qah} and polarization~\cite{Sato-Polito:2023spo,Ellis:2023owy}. Furthermore, the distribution of realizations shows substantial deviations from Gaussianity~\cite{Ellis:2023owy, Ellis:2023dgf, Allen:2024rqk, Lamb:2024gbh, Sato-Polito:2024lew, Xue:2024qtx, Lamb:2025niq}. These are characterized most conveniently by the SMBH GW amplitude distribution (GWAD).

Building on our earlier results~\cite{Ellis:2023owy, Ellis:2023dgf}, we confirm that the high-amplitude tail of GWAD exhibits a \emph{universal}, model-independent power-law scaling $\propto A^{-4}$. This behavior arises from the possibility of having nearby sources and was previously derived for the strain distribution of compact binary backgrounds in~\cite{Ginat:2019aed,Ginat:2023eto}. By contrast, a Gaussian tail, which could be expected from the central limit theorem, would be exponentially suppressed. Consequently, SMBH binary populations dominated by a few loud binaries are relatively likely. More importantly, the distribution of timing residuals $|\delta t_k|$ inherits this behavior, leading to divergent moments of order $n \geq 3$, that is, $\langle |\delta t_k|^n \rangle \to \infty$ with the average taken over different realizations of SMBH populations.

We further demonstrate that the low-amplitude regime of GWAD encodes both the mass dependence of the merger rate and the binary’s energy dissipation mechanisms. Assuming circular, GW-driven binaries and a power-law merger rate $\td R/\td \mathcal{M} \propto \mathcal{M}^{\zeta}$ at low chirp masses, the low-amplitude tail of GWAD follows power-law scaling $\propto A^{-(7-3\zeta)/5}$. This regime is also directly reflected in the distribution of timing residuals, particularly at high frequencies where the number of contributing binaries is smaller.

One must distinguish between the fluctuations of timing residuals within a single SMBH binary population and the variability across different realizations of such populations. Although we only observe one realization and cannot directly access this ensemble variation, it should nonetheless be incorporated into the PTA likelihood, as it affects the statistical inference. Quantifying these non-Gaussian effects is therefore essential both to assess the validity of standard Gaussian PTA analyzes and to motivate frameworks that consistently capture the non-Gaussianity induced by the GWAD. Ways of testing non-Gaussianity in the PTA data have been proposed in~\cite{Lentati:2014hja,Bernardo:2024uiq,Falxa:2025qxr,Lamb:2025niq,Kuntz:2026usl}.

Current PTA analyses~\cite{NANOGrav:2023gor, EPTA:2023fyk, Reardon:2023gzh, Xu:2023wog} assume Gaussian statistics, but the GW background from SMBH binaries is intrinsically non-Gaussian. Properly accounting for this non-Gaussianity is crucial for unbiased SMBH model inference. We confirm that the variance-averaged Gaussian approximation, suggested in~\cite{Xue:2024qtx}, provides an accurate approximation of the timing residual statistics. This supports a factored likelihood framework developed in~\cite{Ellis:2023dgf}, that combines the Gaussian-process PTA posteriors with a non-Gaussian population prior, allowing consistent incorporation of non-Gaussian effects into SMBH model analyses.

We provide a public code \texttt{GWADpy}~\cite{Raidal_GWAD_2026} to compute GWAD from a given SMBH merger rate and to evaluate the corresponding distribution of total PTA timing residuals for a given GWAD. Going beyond our previous studies~\cite{Ellis:2023owy, Ellis:2023dgf}, the code includes interference terms and incorporates window functions that separate sources into different Fourier modes beyond the top-hat approximation. We also investigate the impact of data-processing effects on the window function and on the correlations between Fourier modes. Our numerical implementation is efficient, leveraging the separation between strong and weak sources and an analytical treatment of the high timing residual tail, as in our previous work~\cite{Ellis:2023owy, Ellis:2023dgf}. The code is also flexible, allowing users to input different SMBH merger rates or GWADs, as well as alternative window functions.

This paper is organized as follows. In Sec.~\ref{sec:timing_residuals}, we begin with a brief overview of how GWs from SMBHs give rise to the pulsar timing residuals. Sec.~\ref{sec:GWAD} introduces the GWAD and details its characteristic power-law shape at low and high amplitudes. The distribution of timing residuals induced by the GWAD is derived in Sec.~\ref{sec:stat_timing_residuals}, and the results are discussed in Sec.~\ref{sec:discussion}. We conclude in Sec.~\ref{sec:concl}. Technical details related to the PTA response and the window functions are given in Appendices~\ref{app:response} and \ref{app:window_f}.

\section{Timing residuals}
\label{sec:timing_residuals}

The metric perturbation induced by the GWs emitted by $N$ inspiralling SMBH binaries can be expressed as
\be
	h_{ab}(t,\vec{x}) = \sum_{j=1}^N \sum_{\lambda=+,\times} h_j^\lambda(t - \hat{k}_j\cdot \vec{x}) e_{ab}^\lambda(\hat{k}_j,\psi_j)\,,
\ee
where $j$ labels the binaries, $-\hat{k}_j$ denotes their sky locations, $\psi_j$ denotes their polarization angles, and $e_{ab}^\lambda(\hat{k}_j,\psi_j)$ are the polarization tensors. The polarization modes $h_j^\lambda(t)$ are
\bea\label{eq:pol_modes}
	&h_j^+(t) = \frac{1+\cos^2\imath_j}{2} \, A_j \cos(2\pi f_j t + \delta_j)\,, \\
	&h_j^\times(t) = \cos\imath_j \, A_j \sin(2\pi f_j t + \delta_j)\,,
\eea
where $f_j$ denotes the GW frequency, $\imath_j$ the binary inclination, and $\delta_j$ the phase of the signal. The GW amplitude $A_j$ from a binary with chirp mass $\mathcal{M}_j$ at luminosity distance $D_{L,j}$ is\footnote{We use geometric units with $c = G = 1$.}
\be \label{eq:A1}
    A_j \equiv \frac{4 (1+z_j) \mathcal{M}_j^{\frac53} (2\pi f_{b,j})^{\frac23}}{D_{L,j}} \,,
\ee
where $f_{b,j} = (1+z_j) f_j/2$ is the binary orbital frequency. Note that this only holds for a circular binary. 

PTAs monitor dozens of millisecond pulsars at distances $L_J \sim {\rm kpc}$ from the Earth. The response of a PTA to GWs is encoded in the timing residual, which, for a pulsar located in the direction $\hat{u}_J$ at a distance $L_J$, observed at time $t$, is
\bea \label{eq:deltatJ}
	\delta t_J(t) 
    &= \frac{\hat{u}_J^a \hat{u}_J^b}{2} \int_0^{L_J} \td s \, h_{ab}(t(s), \vec{x}(s)) \\
    &= \sum_{j=1}^N \frac{A_j}{4\pi i f_j} R_{J,j} \,e^{i(2\pi f_j t + \delta_j)} + {\rm c.c.} \,,
\eea
where $t(s) = t - (L_J-s)$ and $\vec{x}(s) = (L_J-s) \hat{u}_J$ parametrize the path from the pulsar to the Earth. The response function is given by
\bea\label{eq:response}
    R_{J,j} 
    &\!=\! \left[ 1\!-\!e^{-2\pi i f_j L_J (1+\hat{k}_j\cdot \hat{u}_J)} \right] \\
    &\,\,\, \times\!\left[\frac{1\!+\!\cos^2 \imath_j}{2} F_J^+(\hat{k}_j,\psi_j)\!-\!i \cos \imath_j F_J^\times(\hat{k}_j,\psi_j) \right]\!,
\eea
with the antenna pattern functions
\be
    F_J^\lambda(\hat{k}_j,\psi_j) 
    = \frac12 \frac{\hat{u}_J^a \hat{u}_J^b}{1+\hat{k}_j\cdot\hat{u}_J} e_{ab}^\lambda(\hat{k}_j,\psi_j) \,.
\ee

We decompose the timing residuals into a discrete Fourier series with frequencies $f_k \equiv k/T$, where $T$ denotes the observation time. The Fourier coefficients are given by
\be \label{eq:deltat}
    \tilde{\delta t}_{J,k} 
    = \sum_{j=1}^N \frac{A_j |R_{J,j}|}{4\pi i f_j} \bigg[ e^{i\bar\delta_{J,j}} w_{k,j}^+ - e^{-i\bar\delta_{J,j}} w_{k,j}^- \bigg] \,,
\ee
where $\bar \delta_{J,j} \equiv \delta_j + \arg R_{J,j}$ and $w_{k,j}^\pm = w_{k}(\pm f_j)$ are window functions. 

Direct computation of the Fourier coefficients from Eq.~\eqref{eq:deltatJ} gives $w_k(f) = \sinc[\pi T (f - f_k)]$, leading to strong leakage between Fourier modes. This can be mitigated in data processing, for example, through pre-whitening and post-coloring~\cite{Coles:2011zs}. Furthermore, PTA data analysis subtracts noise/background components that cannot be distinguished from the signal~\cite{Taylor:2021yjx}. Both of these procedures modify the window function. We discuss these modifications and their impact in Appendix~\ref{app:window_f}. 

An idealized band-pass filter corresponds to a top-hat window function equal to 1 when $|f_j-f_k| < 1/(2T)$ and 0 otherwise. While this cannot be realized in finite time measurements due to the uncertainty between frequency and time, we use it to represent an idealized case that approximates situations where spectral leakage into neighboring Fourier modes, and thus the correlations induced between modes, can be efficiently suppressed by data processing. We note that correlations can still arise when the sources are not monochromatic. This occurs, for example, if SMBH binaries are eccentric, although near maximal eccentricity is required to produce significant correlations~\cite{Raidal:2024odr}.

Assuming that the binary sky location, inclination, polarization, and phase are independent and uniformly distributed, the quantity $e^{i\delta_j} R_{J,j} \equiv e^{i\bar\delta_{J,j}} |R_{J,j}|$ is a complex random variable with a uniformly distributed phase $\bar\delta_{J,j} \in [0,2\pi)$, independent of its modulus, which lies in the range $[0,2]$ (see Appendix~\ref{app:response}). The properties of the binary population are encoded in the GW amplitudes $A_j$ and the frequencies $f_j$. We discuss their statistical properties in Sec.~\ref{sec:GWAD}.

We have neglected the evolution of the binary frequency. As discussed in Appendix~\ref{app:response}, the frequency change enters only through the pulsar-term phase in the response function $R_{J,j}$, Eq.~\eqref{eq:response}, and it can be neglected as long as $\dot f_j L_J T \ll 1$. A circular GW driven binary evolves as $\dot f_j/f_j = 3 t_{\rm GW}^{-1}/2$, where the characteristic inspiral timescale is
\bea \label{eq:tcirc}
    t_{\rm GW} &= \frac{5}{64} \frac{1+z}{\mathcal{M}^{5/3} (2 \pi f_b)^{8/3}} \\
    &\approx 28 \,\mathrm{kyr} \,(1+z) \left[\frac{\mathcal{M}}{10^{9}\,\Msun}\right]^{-\frac53} \left[\frac{f_b}{10\,\mathrm{nHz}}\right]^{-\frac83} \,.
\eea
Therefore, since $t_{\rm GW}$ exceeds the observation time $T \sim 10\,$yr by three orders of magnitude, each source remains monochromatic to high accuracy over the course of the observations.
Furthermore, for most binaries contributing at nHz frequencies, the gravitational-wave timescale $t_{\rm GW}$ is roughly an order of magnitude larger than the light-travel time between a given pulsar and the Earth, $L_J = 3.3\,\mathrm{kyr}\,(L_J/{\rm kpc})$. Consequently,
\be
    \dot f_j L_J T \approx 0.13 \!\left[\frac{(1+z)\mathcal{M}}{10^{9}\,\Msun}\right]^{\frac53} \left[\frac{f_j}{10\,{\rm nHz}}\right]^{\frac{11}{3}} \frac{L_J}{\rm kpc} \frac{T}{15\,{\rm yr}} \,,
\ee
indicating that the change in frequency between the Earth term and the pulsar term can be safely neglected in the lowest frequency bins. For frequencies above 20\,nHz, the pulsar-term contribution can move into adjacent frequency bins. Since this contribution enters only through the pulsar term, its impact is expected to remain small~\cite{Ellis:2012zv}, and a detailed analysis is deferred to future work.

\section{GW amplitude distribution}
\label{sec:GWAD}

\subsection{Definition}

The GW amplitude distribution (GWAD) is the distribution of GW amplitudes $A$ from individual binaries at a given frequency $f$. It is the central object underlying the statistical properties of the SMBH GW power spectrum and the induced timing residuals, and it is given by
\be\label{eq:P1semi-analytic}
    \frac{\td N}{\td A \,\td \ln f} 
    = \int \td\lambda \frac{\td t}{\td\ln{f_{\rm b}}} \delta(A - A^{(1)}) \big|_{f_{\rm b} = \frac{(1+z)f}{2}} \,,
\ee
where $\td \lambda$ is the differential merger rate of BHs in the observer reference frame, $\td t/\td \ln{f_{\rm b}}$ is the residence time of the binary, and $A^{(1)} \equiv A^{(1)}(f_{b},\mathcal{M},D_L)$ denotes the GW amplitude from a single circular binary (see Eq.~\eqref{eq:A1}). 

At sufficiently low frequencies (large separations), the binary evolution is driven by its interactions with surrounding gas and stars~\cite{Armitage:2002uu,2013CQGra..30x4005M,Kelley:2016gse,Tang:2017eiz}. We incorporate these environmental effects through a characteristic timescale $t_{\rm env}$, which modifies the binary frequency evolution as
\be
     \frac{\td t}{\td \ln f_b} = \frac23 \frac{1}{t_{\rm GW}^{-1} + t_{\rm env}^{-1}} \,.
\ee
For a circular binary the GW timescale is given in Eq.~\eqref{eq:tcirc}, while the environmental contribution is parametrized as~\cite{Ellis:2023dgf}
\be
    t_{\rm env} = t_{\rm GW} \bigg[ \frac{2f_b}{f_{\rm ref}(\mathcal{M}/10^9\Msun)^\beta} \bigg]^\alpha \,,
\ee
where $f_{\rm ref}$ sets the transition frequency below which environmental effects dominate, and $\alpha$ and $\beta$ control the scaling with frequency and chirp mass.

\begin{figure*}
    \centering
    \includegraphics[width=\linewidth]{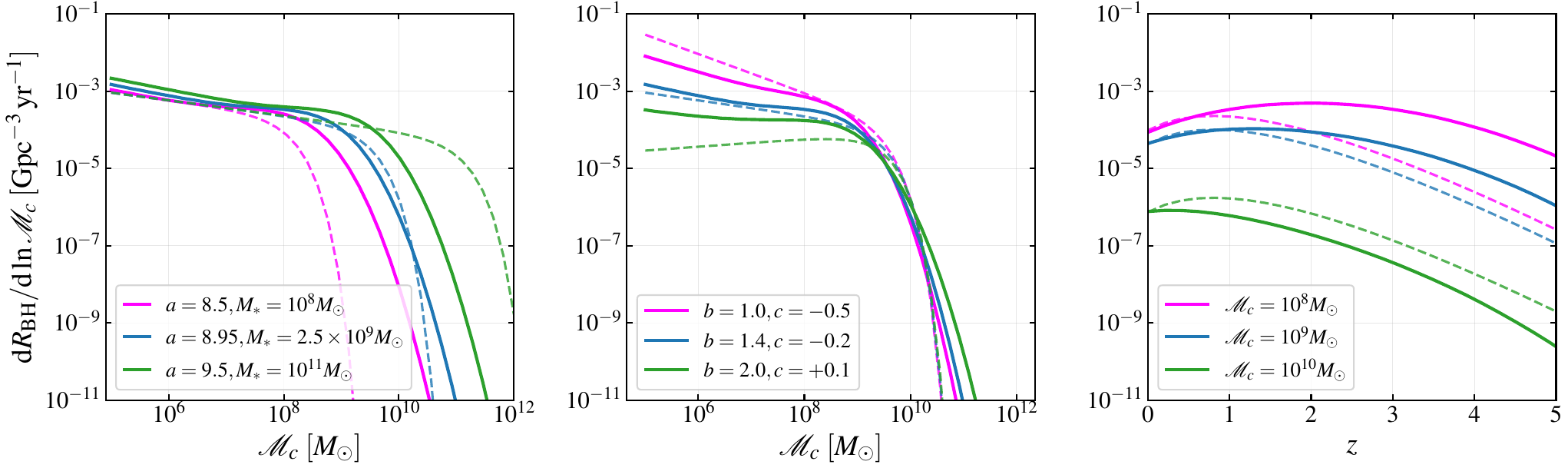}
    \caption{The SMBH merger rates of Model I~\eqref{eq:modelA} (solid) and Model II~\eqref{eq:modelB} (dashed). The parameters that are not varied are fixed to the fiducial values $\{p_{\rm BH},a,b,\sigma\}=\{0.6,8.95,1.4,0.47\}$ for Model I and $\{R_0,c,d,z_0,\mathcal{M}_*\}=\{4\times10^{-5}\,{\rm Gpc}^{-3}{\rm yr}^{-1},-0.2,6.0,0.3,2.5\times10^{9}\,\Msun\}$ for Model II. The left and middle panels show the merger rate at $z = 1$.}
    \label{fig:merger_rate}
\end{figure*}

The properties of the SMBH binary population are determined by their comoving merger rate density $\td R_{\rm BH}/\td \mathcal{M}\td \eta$ that enters through $\td\lambda$:
\be \label{eq:diffmergerrate}
    \td \lambda = \td \mathcal{M} \td \eta \td z \frac{1}{1+z} \frac{\td V_c}{\td z} \frac{\td R_{\rm BH}}{\td \mathcal{M} \td \eta}\,.
\ee
Here $\eta$ is the symmetric mass ratio of the binary, and $V_c$ is the comoving volume, whose derivative in terms of luminosity distance $D_L$ and the Hubble rate $H$ is
\be
    \frac{\td V_c}{\td z} = \frac{4\pi}{H} \frac{D_L^2}{(1+z)^2} \,.
\ee 
In order to evaluate the GWAD, we consider two models for the SMBH merger rate density:

In \textbf{Model I}, the BH merger rate is obtained from the halo merger rate $R_h$:
\be \label{eq:modelA}
    \frac{\td R_{\rm BH}}{\td m_1 \td m_2} \!=\! p_{\rm BH} \int \! \td M_1 \td M_2 \frac{\td R_h}{\td M_1 \td M_2} \prod_{j=1,2} \!\frac{\td P(m_j|M_j)}{\td m_j}  , 
\ee
where $m_{j}$ are the masses of the merging BHs, $M_j$ are the masses of their host halos and $p_{\rm BH} \le 1$ combines the SMBH occupation fraction in galaxies with the efficiency for the BHs to merge following the merger of their host halos. We use the halo merger rate arising from the extended Press-Schechter formalism~\citep{Press:1973iz,Bond:1990iw,1993MNRAS.262..627L}, relate the halo masses to the galaxy stellar masses using the fit of Ref.~\cite{Girelli:2020goz}, and parametrize the BH mass-stellar mass relation as
\be \label{eq:Ms_relation}
    \frac{\td P(m|M_*)}{\td \log_{10} \!m} = \mathcal{N}\bigg(\!\log_{10} \!\frac{m}{M_\odot} \bigg| a + b \log_{10} \!\frac{M_*}{10^{11}M_\odot},\sigma\bigg) \, ,
\ee
where $\mathcal{N}(x|\bar x,\sigma)$ denotes the probability density function (PDF) of a Gaussian distribution with mean $\bar x$ and variance $\sigma^2$. Fits of the local dynamically detected SMBHs, corresponding to the heaviest local SMBH population, give $a = 8.95$, $b = 1.4$, and $\sigma = 0.47$~\cite{Reines:2015nyy}, which we take these as the fiducial values. Furthermore, we fix $p_{\rm BH} = 0.6$.

In \textbf{Model II}, the BH merger rate is parametrized, following Ref.~\cite{Middleton:2015oda}, as a distribution of chirp mass and redshift, with the dependence on $\eta$ integrated out. It features a power-law behavior in $\mathcal{M}$ for $\mathcal{M}\ll\mathcal{M}_*$ with an exponential cut-off near $\mathcal{M} \simeq \mathcal{M}_*$, as well as a low-$z$ power-law scaling in $1+z$ with an exponential cut-off around $z \simeq z_0$:
\be \label{eq:modelB}
    \frac{\td R_{\rm BH}}{\td \mathcal{M}} = \frac{R_0}{\mathcal{M}} \left(\frac{\mathcal{M}}{10^{10} \Msun}\right)^{c} e^{-\mathcal{M}/\mathcal{M}_*} (1+z)^d e^{-z/z_0} \,.
\ee
As fiducial values in Model~II, we adopt $\mathcal{M}_* = 2.5 \times 10^9\,\Msun$ and $c = -0.2$, which yield a chirp mass dependence similar to that of the fiducial Model~I, and $R_0 = 4\times10^{-5}\,{\rm Gpc}^{-3}{\rm yr}^{-1}$, which matches the fiducial Model~I amplitude at $z\approx 0$. Furthermore, we fix $d = 6$ and $z_0 = 0.3$.

The solid and dashed curves in Fig.~\ref{fig:merger_rate} show the BH merger rates of Model I and Model II, respectively. The chirp mass dependence in both models exhibits a power-law tail at low masses and an exponential cut-off at high masses. In Model~I these features are inherited from the halo merger rate and the low-mass slope and the position of the exponential cut-off depend on the BH mass-stellar mass relation~\eqref{eq:Ms_relation}: the low-mass power-law depends on $b$ (the exponent in the power-law relation between halo mass and BH mass), while the position of the high-mass cut-off is set by $a$ (the proportionality constant in the halo mass–BH mass relation). Accordingly, the parameters $a$ and $b$ in Model I play roles analogous to those of $\mathcal{M}_*$ and $c$ in Model II, respectively. This is illustrated in the left and middle panels of Fig.~\ref{fig:merger_rate}. The right panel of Fig.~\ref{fig:merger_rate} shows that the redshift evolution of the merger rate in Models I and II is qualitatively similar.

\begin{figure*}
    \centering
    \includegraphics[height=0.29\textwidth]{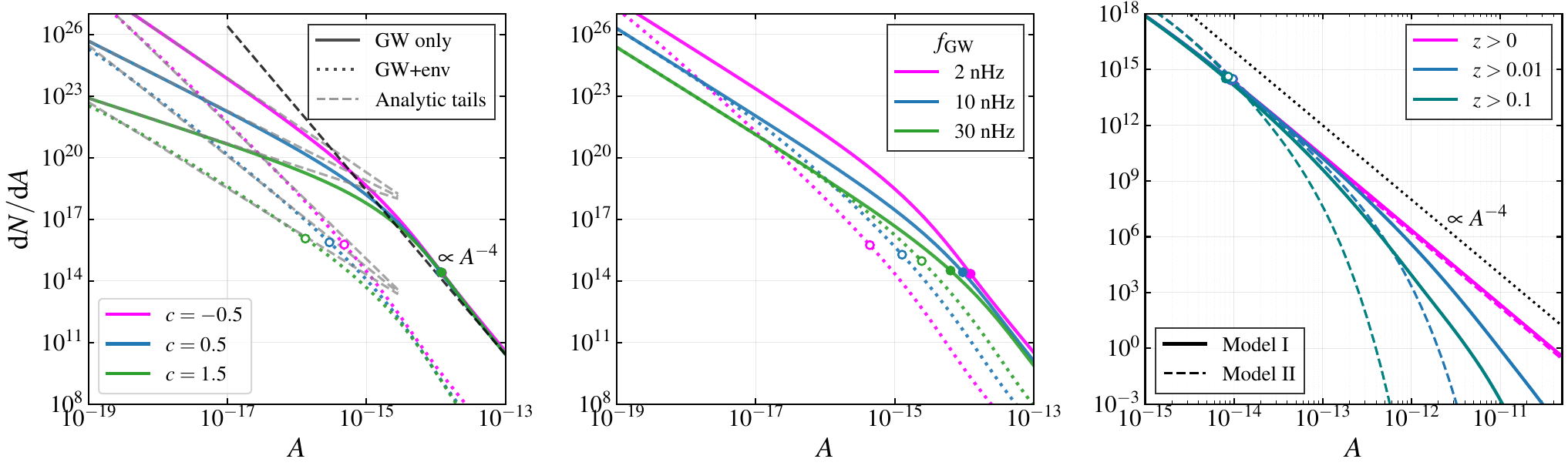} 
    \caption{The solid and dotted curves in the left and middle panels show, respectively, the GWAD for binaries evolving purely via GW emission and for binaries affected by environmental effects with $f_{\rm ref} = 30\,$nHz, $\alpha = 8/3$, and $\beta = 5/8$ in merger rate Model II. The \emph{left panel} shows the GWAD integrated over the frequency range $(1,3)\,$nHz for different values of the Model II parameter $c$ that determines the low-mass power-law, while in the \emph{middle panel} $c = -0.2$ and the GWAD is shown for frequency ranges $(1,3)\,$nHz, $(9,11)\,$nHz, and $(29,31)\,$nHz. The dot on each line corresponds to the threshold amplitude above which only one binary is expected. In the \emph{right panel}, the solid and dashed curves show the GWAD in Model I and Model II, respectively, integrated over the frequency range $(1,3)\,$nHz for different redshift cuts applied to the SMBH binary population. The merger rate parameters are the same as those used in Fig.~\ref{fig:merger_rate}.}
    \label{fig:tails}
\end{figure*}

\subsection{Properties of GWAD}

Regardless of the physical properties of the SMBH binary population, such as orbital eccentricity, environmental interactions, or binary masses, GWAD exhibits a universal broken power-law shape. Below, we explain the physical origin of the two asymptotic power-law regimes.

\subsubsection{High-amplitude tail}
\label{sec:high-A}

The high-$A$ asymptotic of GWAD is independent of the modeling of the SMBH merger rate. It originates from the possibility of having nearby sources and can be derived analytically by considering potential $z\ll 1$ binaries. The probability of finding such a nearby binary is proportional to the area of the shell around the observer, $\td \lambda \propto D_L^2 \td D_L$, and GW emission by such a binary produces $A^{(1)}\propto 1/D_L$ (see Eq.~\eqref{eq:A1}). The high-$A$ asymptotic is obtained by considering very nearby, $z\sim 0$, sources:
\begin{align}\label{eq:A-4_tail}
    \frac{\td N}{\td A \,\td \ln f} &\stackrel{z\sim 0}{=} \int \td\mathcal{M} \td\eta \frac{\td R_{\rm BH}}{\td\mathcal{M} \td\eta} \frac{\td t}{\td \ln f_b} \nonumber\\
    &\quad \times 4\pi \int \td D_L \, D_L^2 \,\delta(A-A^{(1)}) \big|_{z = 0} \\
    & = C_{\infty} \, A^{-4}  \,,\nonumber
\end{align}
where the normalization GWAD's heavy power-law tail is
\be
    C_{\infty} = 256\pi^3 f^2 \!\int\! \td\mathcal{M} \td\eta \, \mathcal{M}^5 \frac{\td R_{\rm BH}}{\td\mathcal{M} \td\eta} \frac{\td t}{\td \ln f_b}  \bigg|_{z = 0}.
\ee
For binaries evolving due to GW emission, the normalization factor scales with frequency as $C_{\infty} \propto f^{-2/3}$, whereas for binaries whose evolution is driven by environmental effects, the scaling becomes $C_{\infty} \propto f^{-2/3 + \alpha}$. The $\mathcal{M}^5$ dependence in the integrand of the normalization factor indicates that the high-$A$ tail is dictated by the heavy end of the binary population.

It is important to stress that the $A^{-4}$ asymptotic power-law is universal, as it arises simply from the possibility of having an arbitrarily nearby binary and is therefore not dependent on the choice of the merger rate or the SMBH binary population. This is illustrated in Fig.~\ref{fig:tails}, where we see that varying the merger rate and environmental interactions only affects the low-amplitude asymptotic power, while the high-amplitude tail remains fixed. This is further demonstrated in the right panel of Fig.~\ref{fig:tails} by applying redshift cuts to the binary population. Imposing a sufficiently large minimal redshift for potential SMBH binaries removes the power-law tail. Nevertheless, in the range of amplitudes most relevant to the current PTA experiments,\footnote{Searches of individual SMBH binaries in the PTA data have excluded amplitudes $A > 10^{-14}$ in the $1-100\,{\rm nHz}$ frequency range~\cite{NANOGrav:2023pdq}.} $A \lesssim 10^{-14}$, a power law behavior of the GWAD is generally still retained. For this reason, imposing such cuts does not significantly alter our conclusions.

The dots in Fig.~\ref{fig:tails} show the amplitude $A_c$ above which the expected number of sources is one or less, that is, 
\be
    \int_{A_c}^\infty \td A \frac{\td N}{\td A} = 1 \,. 
\ee
If $A_c$ lies on the $A^{-4}$ tail, then $A_c = \sqrt[3]{C_{\infty}/3}$. However, as seen from Fig.~\ref{fig:tails}, this is not always the case. In particular, the single source regime starts at a lower value of $A$ at higher frequencies because the binaries evolve faster, and consequently, their number decreases with increasing $f$. The same effect is observed at low frequencies in the case of environmental effects that make the evolution of binaries faster.

\subsubsection{Low-amplitude tail}

The emergence of the low-$A$ power-law can clearly be seen in Fig.~\ref{fig:tails}, where we also show that the tail varies when the parameter determining the low-$\mathcal{M}$ power-law of the merger rate is varied. This suggests that the low-amplitude tail of the GWAD is largely governed by the low-mass binary population. A merger rate that follows a power-law in the chirp mass, $\td R_{\rm BH}/\td\mathcal{M}\td\eta \propto \mathcal{M}^\zeta$, yields a GWAD that is a power-law:
\bea\label{eq:GWAD-lowA}
    \frac{\td N}{\td A \,\td \ln f} 
    &= \int \td\eta \frac{\td z}{1+z} \frac{\td V_c}{\td z} \frac{\td R_{\rm BH}}{\td\mathcal{M}\td\eta} \bigg|_{\mathcal{M} = \mathcal{M}_0} \\
    &\quad\times 
    \frac{3 \mathcal{M}_0}{5 A} \left[\frac{\mathcal{M}}{\mathcal{M}_0}\right]^{\zeta+1} 
    \frac{\td t}{\td \ln f_b}  \bigg|_{\mathcal{M}: A = A^{(1)}} \\
    &\propto f^{-\frac{12}5 -\frac25 (\zeta-\alpha\beta) + \alpha} A^{-\frac75 + \frac35 (\zeta-\alpha\beta)} \,,
\eea 
where the GW driven case corresponds to $\alpha = 0$. Note that in Model II, we have $\zeta = c-1$. Although we do not give the normalization, the scaling property of $\td N/\td A \,\td \ln f$ allows to estimate it easily by computing the integral~\eqref{eq:GWAD-lowA} at some reference values $A = A_{\rm ref}$ and $f = f_{\rm ref}$.

As seen in the left panel of Fig.~\ref{fig:tails}, Eq.~\eqref{eq:GWAD-lowA} accurately captures the low-$A$ behavior of the GWAD with $\zeta$ determined from the low mass power-law scaling of the SMBH merger rate. Unlike the universal high-$A$ scaling of $A^{-4}$, the low-$A$ power depends on the merger rate parameters and the energy loss mechanisms (e.g., environmental effects) of the SMBH binaries.

\section{Statistics of the timing residuals}
\label{sec:stat_timing_residuals}

As shown in Eq.~\eqref{eq:deltat}, the timing residuals induced by the GWs emitted by a population of SMBH binaries can be expressed as a sum of a large number of independent identical random variables. Although the variance of the amplitudes is finite and the central limit theorem formally applies, the presence of a heavy tail $\propto A^{-4}$ implies that moments of order three and higher diverge. As a result, the convergence to Gaussian behavior is slow with an increasing number of sources, and while the peak is approximately Gaussian, the distribution of timing residuals retains significant non-Gaussian features even if the number of GW sources is large in each realization of the SMBH population. To quantify these non-Gaussian effects, we compute the probability distribution of the magnitude of the timing residual Fourier coefficients $|\tilde{\delta t}_{J,k}|$.

\subsection{Gaussian approximation}

Let us begin by characterizing the Gaussian approximation that is often adopted in the literature. In this approximation, the distribution of $\tilde{\delta t}_{J,k}$ is determined solely by the covariance matrix of the Fourier modes,
\bea \label{eq:corr_dt}
    \langle \tilde{\delta t}_{k} \tilde{\delta t}^{*}_{k'} \rangle 
    &= \frac{\bar{N}}{60\pi^2} \, \bigg\langle \frac{A^2}{f^2}
    \bigg[w_{k}^+ w_{k'}^+ + w_{k}^- w_{k'}^-  \bigg]
    \bigg\rangle_{\!A,f} \,,
\eea
where we averaged over phases, polarizations, inclinations, and sky locations, which yields $\langle|R|^2\rangle = 4/15$ in the $f L \gg 1$ limit (see Appendix~\ref{app:response}), and $\bar{N}$ denotes the expected number of binaries. The remaining average is over the amplitudes $A$ and the frequencies $f$. We have also dropped the indices $J$ and $j$, since all sources are statistically identical, and the sum over sources is accounted for by the prefactor $\bar{N}$. 

Although leakage between Fourier modes cannot be completely avoided by data processing techniques, it can be sufficiently suppressed, and the top-hat window function can be adopted as an idealized approximation (see Appendix~\ref{app:window_f}). This yields the diagonal covariance matrix
\bea\label{eq:dt_cov}
    \langle \tilde{\delta t}_{k} \tilde{\delta t}^{*}_{k'} \rangle
    &\approx \frac{\delta_{kk'}}{60 \pi^2 T f_k^2} \int \td A \, A^2 \frac{\td N}{\td f \td A} \bigg|_{f = f_k}
    \,.
\eea
For dominantly GW driven binaries, the integral in~\eqref{eq:dt_cov} scales as $\propto f_k^{-7/3}$ (see Eqs.~(\ref{eq:P1semi-analytic}~-~\ref{eq:tcirc})), and we obtain 
\be
    \langle |\tilde{\delta t}_{k}|^2 \rangle \propto f_k^{-13/3} \,.
\ee

\begin{figure*}
  \centering
  \includegraphics[width=\linewidth]{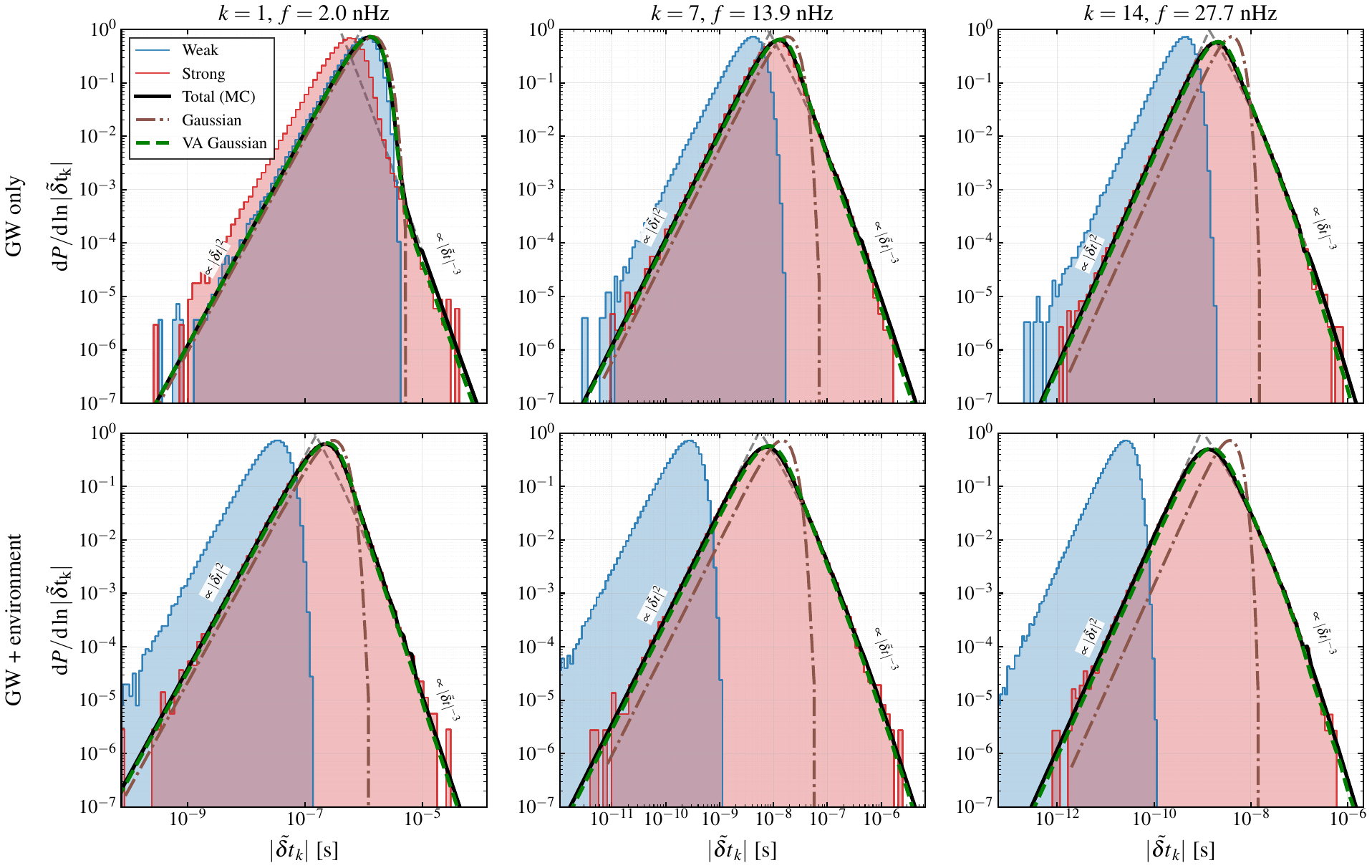}
  \caption{The distribution of timing residuals generated with \texttt{GWADpy}. Shown are the MCMC histograms for weak and strong source timing residuals at three different frequencies, as well as the total obtained through attaching the low and large residual analytical tails. Shown as a dot-dashed brown curve is the distribution in the Gaussian approximation~\eqref{eq:corr_dt} and as a dashed green curve is the distribution in the variance-averaged (VA) Gaussian approximation~\eqref{eq:VA_Gaussian}. The distributions are obtained from $10^6$ realizations, with $N_{\rm S}=50$ and $N_{\rm bins}=200$ using the top-hat window function. The source population is derived using Model I merger rate with $\{p_{\rm BH},a,b,\sigma\}=\{0.6,8.95,1.4,0.47\}$ and environmental effects (in the bottom row) with $\{\alpha,\beta,f_{\rm ref}\}=\{8/3,5/8,30\,\rm{nHz}\}$.}
  \label{fig:histograms}
\end{figure*}

In the Gaussian approximation, the covariance matrix~\eqref{eq:corr_dt} provides a complete statistical description of the timing residuals. Since the Fourier coefficients $\tilde{\delta t}_{J,k}$ are complex, they follow a bivariate Gaussian distribution, and their absolute values follow a Rayleigh distribution,
\be\label{eq:dt_Gaussian}
    \frac{\mathrm{d} P}{\mathrm{d} \ln|\tilde{\delta t}_k|}
    = \frac{|\tilde{\delta t}_k|^2}{\sigma_k^2}
      \exp\!\left[-\frac{|\tilde{\delta t}_k|^2}{2\sigma_k^2}\right] \,,
\ee
where $\sigma_k^2 \equiv \langle |\tilde{\delta t}_{k}|^2 \rangle$. This approximation is shown in Fig.~\ref{fig:histograms} by the brown dashed curves. Compared to the true distribution shown in black, we see that the Gaussian approximation fails badly when the signal is dominated by a handful of loud binaries. This occurs at high frequencies and in the presence of strong environmental effects. In Fig.~\ref{fig:histograms}, reasonable agreement with the Gaussian approximation at the peak can be observed only in the upper left panel, corresponding to the lowest $2$\,nHz frequency mode and purely GW-driven binaries.

\subsection{Non-Gaussian statistics}

It is known that the SMBH GW background exhibits non-Gaussian features that manifest themselves as heavy power-law tails~\cite{Ellis:2023owy,Ellis:2023dgf}. To accurately quantify such features, we will evaluate the distribution of timing residuals for a single pulsar numerically. 

The number of sources contributing to any given mode can be enormous, which makes a direct Monte Carlo sum over all of them, computed via~\eqref{eq:deltat}, prohibitively expensive. To overcome this issue, we follow~\cite{Ellis:2023owy} and split the sources into strong and weak sources by threshold amplitude $A_{\rm th}$. This splits the total timing residual into two components
\be\label{eq:sum}
    \tilde{\delta t}_{J,k} 
    = \tilde{\delta t}^{\rm S}_{J,k} +
    \tilde{\delta t}^{\rm W}_{J,k} \,,
\ee
with S and W labeling the strong ($A^{\rm S} \geq A_{\rm th}$) and weak ($A^{\rm W} < A_{\rm th}$) contributions, respectively. This split cuts off the heavy tail in the distribution of the weak sources. In particular, the threshold can be chosen so that the weak component is approximately Gaussian\footnote{This requires $A_{\rm th}$ to be sufficiently low. When choosing $A_{\rm th}$, we have checked that the total distribution remains unchanged when computed with an even lower $A_{\rm th}$.}, while the less numerous strong sources will be responsible for the non-Gaussian features. This will greatly reduce the number of terms in the sum in Eq.~\eqref{eq:deltat} and significantly speed up the generation of the timing residual PDFs.

\subsubsection{Strong sources}

To sample the contributions of the timing residuals, we divide a broad frequency band into a large number $N_{\rm bins}$ of narrow bins in binary frequency. The width of the band is conditional on the amount of leakage allowed by the window function in Eq.~\eqref{eq:deltat}. For a given Fourier mode at $f_k$, the amplitudes of the $N_{\rm S}$ strong sources are sampled explicitly from the GWAD, conditional on $A \geq A_{\rm th}$ in each binary frequency bin. The contribution of all the strong sources to the total timing residual is then computed as
\be \label{eq:deltatS}
    \!\tilde{\delta t}_{J,k}^{\rm S} 
    \!=\!\!\sum_{j=1}^{N_{\rm bins}} \!\sum_{j'=1}^{N_{\rm S}} \frac{A_{j'} |R_{J,j'}|}{4\pi i f_j} \bigg[ e^{i\bar\delta_{J,j'}} w_{k,j}^+\!- e^{-i\bar\delta_{J,j'}} w_{k,j}^- \bigg],
\ee
where the first sum runs over the binary frequency bins and the second over the sampled strong sources that bin. The amplitudes $A_{j'}$ are sampled from GWAD integrated over the frequency bin $j$, while $\bar\delta_{J,j'}$ is drawn from a uniform distribution on $[0,2\pi)$, and $|R_{J,j'}|$ is sampled from the PDF given in Appendix~\ref{app:response}.

The contribution of strong sources, calculated with $N_S = 50$ and $N_{\rm bins} = 200$ by generating $10^5$ realizations of $|\tilde{\delta t}_{J,k}^{\rm S}|$ and binning them, is shown by the red histograms in Fig.~\ref{fig:histograms}. As expected, it provides the dominant contribution at high $|\tilde{\delta t}_{J,k}|$. It is highly non-Gaussian and exhibits a heavy tail, which we discuss further in Sec.~\ref{sec:tail}.

\subsubsection{Weak sources}

The weak sources are far more numerous but also individually insignificant. Their contribution is included in each binary frequency bin as
\be
    \tilde{\delta t}_{J,k}^{\rm W} 
    = \sum_{j=1}^{N_{\rm bins}} \left[ \mathcal{T}_j w_{k,j}^+ - \mathcal{T}_j^* w_{k,j}^- \right] \,,
\ee
where $\mathcal{T}_j$ is a complex random variable whose statistical moments are all finite, and by the central limit theorem, it is expected to follow a complex Gaussian distribution. This means that, instead of generating the contribution from each binary individually, its possible to sample the total contribution from weak sources by drawing the real and imaginary parts of $\mathcal{T}_j$ independently from a Gaussian distribution with variance 
\bea \label{eq:sigma_jW}
    \sigma^2_{\mathcal{T}_j} 
    &= \frac{\bar N_j}{2} \bigg\langle \frac{A^2 |R|^2}{(4\pi f)^2} \bigg\rangle_{|R|,A,f\in [f_{j,{\rm min}},f_{j,{\rm max}}]} \\
    &= \frac{1}{120 \pi^2} \int_{f_{j,{\rm min}}}^{f_{j,{\rm max}}} \frac{\td f}{f^2} \int_0^{A_{\rm th}} \td A\, A^2 \frac{\td N}{\td f \td A} \,,
\eea
where the frequency integral is over the binary frequency bin $j$. 

It is also possible to draw $\tilde{\delta t}_{J,k}^{\rm W}$ directly from a Gaussian distribution. The benefit of dividing the weak sources into smaller binary frequency bins is that this approach can automatically model correlations between Fourier modes when a realistic window function is considered. The difference in computation time between these approaches is negligible.

The contribution of weak sources is shown by the blue histograms in Fig.~\ref{fig:histograms}, with the threshold $A_{\rm th}$ determined by $N_S = 50$. Similarly to the contribution of strong sources, these histograms are obtained with $N_{\rm bins} = 200$ by generating $10^5$ realizations of $|\tilde{\delta t}_{J,k}^{\rm W}|$ and binning them. We see that this contribution is often negligible compared to that from the strong sources, indicating that $N_S = 50$ is typically more than enough.

The PDF of $|\tilde{\delta t}_{J,k}|$ is then obtained by summing the realizations of strong and weak contributions, as in Eq.~\eqref{eq:sum}. With this setup, the computation itself is rapid, but an impractically large number of samples is required to accurately capture the non-Gaussian, high-amplitude tail caused by the universal high-$A$ behavior of the GWAD, as described in Sec.~\ref{sec:high-A}. This issue can be solved by analytically deriving the tail and attaching it to the simulated distribution.

\subsubsection{Analytic tails}
\label{sec:tail}

Due to the single big jump principle of heavy-tailed distributions (see Sec.~\eqref{sec:SLS}), the statistics of the timing residuals at high $|\tilde{\delta t}_{J,k}|$ are dominated by a single loud source, whose amplitude is described by the high-$A$ tail of the GWAD. The distribution in the high-$|\tilde{\delta t}_{J,k}|$ tail can be estimated by considering the distribution of timing residuals from a single SMBH binary source: 
\begin{widetext}
\bea\label{eq:dt_tail}
	\frac{\td P}{\td |\tilde{\delta t}_k|} 
    \sim \frac{\td N}{\td |\tilde{\delta t}_k|} 
    &= \bar{N} \left\langle \delta\!\left( |\tilde{\delta t}_k| - \frac{A |R| }{4\pi f} |e^{i\bar{\delta}} w_k^+(f) - e^{-i\bar{\delta}} w_k^-(f)| \right)  \right\rangle_{|R|,A,f,\bar{\delta}} 
    \\
    &= \int_0^2 \td |R| \, p(|R|) \int_{-\infty}^\infty \td \ln f  \int_0^{2\pi} \frac{\td \bar\delta}{2\pi} \, \left[ \frac{A}{|\tilde{\delta t}_k|} \frac{\td N}{\td \ln f \td A} \right]_{A = \frac{4\pi f |\tilde{\delta t}_k|}{|R| |e^{i\bar{\delta}} w_k^+(f) - e^{-i\bar{\delta}} w_k^-(f)|}} \,.
\eea
\end{widetext}
As shown in Fig.~\ref{fig:histograms}, the numerically generated PDFs of $|\tilde{\delta t}_{J,k}|$ are in excellent agreement with this distribution at the high-$|\tilde{\delta t}_{J,k}|$ tail.

In general, the integral \eqref{eq:dt_tail} cannot be simplified further. However, at sufficiently large $|\tilde{\delta t}_k|$, it is dominated by the $\propto A^{-4}$ tail of GWAD given in Eq.~\eqref{eq:A-4_tail}. This implies that the distribution of timing residuals then asymptotes to
\be
    \frac{\td P}{\td |\tilde{\delta t}_k|} \sim I_k |\tilde{\delta t}_k|^{-4}\,,
\ee
where the normalization is given by
\bea
    I_k 
    &\approx \frac{1}{256 \pi^3} \int_0^\infty \frac{\td f}{f^4} C_{\infty}(f) \left|w_k^+(f)\right|^{3}\,,
\eea
and we used $\langle |R|^3\rangle = 1/4$. This asymptotic scaling is clearly visible in Fig.~\ref{fig:histograms}, but in particular, in the bottom right panel, incorporating only the $|\tilde{\delta t}_k|^{-4}$ tail is not sufficient, as the total distribution starts to follow the GWAD immediately after the peak. This also means that the distribution near the peak is sensitive to the low-$A$ part of the GWAD and therefore reflects the mass dependence of the merger rate and the binary's energy dissipation mechanisms.

The low-$|\tilde{\delta t}_{J,k}|$ tail arises from destructive interference between signals from different binaries which, in the absence of a single dominant contribution, allows the complex sum to approach zero~\cite{Ginat:2019aed,Xue:2024qtx}. The low-$|\tilde{\delta t}_{J,k}|$ scaling $\td P/\td \ln |\tilde{\delta t}_k| \sim B |\tilde{\delta t}_k|^2$ then follows that of the Rayleigh distribution~\eqref{eq:dt_Gaussian}. We can accurately estimate the normalization $B$ of the low-$|\tilde{\delta t}_{J,k}|$ tail directly from the simulated residuals by matching the cumulative probability
\be
    P(|\tilde{\delta t}_k| < t_{\rm th}) = \int_0^{t_{\rm th}} \td \ln |\tilde{\delta t}_k|\, \frac{\td P}{\td \ln |\tilde{\delta t}_k|} \,,
\ee
where the threshold $t_{\rm th}$ is chosen so that it is well below the peak of the distribution and $P(|\tilde{\delta t}_k| < t_{\rm th})$ is the cumulative probability obtained from the simulated residuals. This gives $B = 2 P(|\tilde{\delta t}_k| < t_{\rm th})/t_{\rm th}^2$. As seen from Fig.~\ref{fig:histograms}, this approach accurately produces the low-residual tail even in the case with strong environmental effects, where the signal is dominated by a small number of strong sources.

\subsection{Variance-averaged Gaussian}
\label{sec:AG}

\begin{figure}[b]
  \centering
  \includegraphics[width=\linewidth]{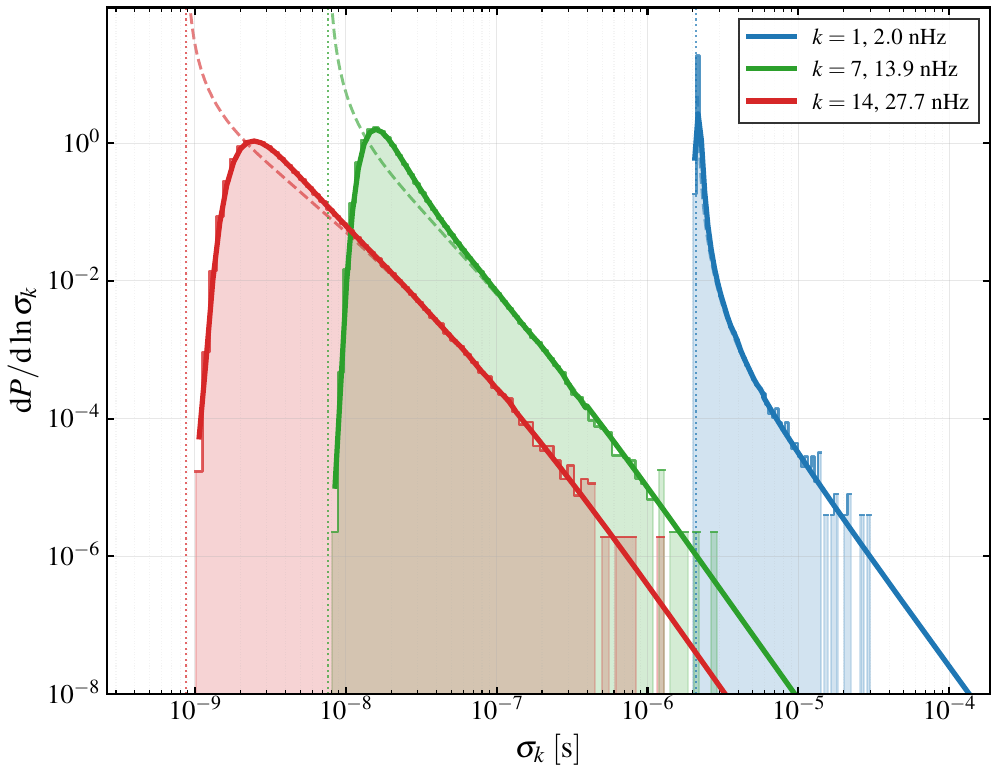}
  \caption{The distribution of the averaged variances~\eqref{eq:sigma0} at three different $k$-modes. The distributions are obtained from $10^6$ realizations, with $N_{\rm S}=50$ and $N_{\rm bins}=200$ using the top-hat window function. The source population is derived using Model I merger rate with $\{p_{\rm BH},a,b,\sigma\}=\{0.6,8.95,1.4,0.47\}$. The vertical dotted line shows the contribution from weak sources and the dashed curve shows the analytic high-$\sigma_k$ tail, computed as in~\cite{Ellis:2023dgf}.}
  \label{fig:sigma0}
\end{figure}

It was demonstrated in~\cite{Xue:2024qtx} that an approximation of the distribution of timing residuals can be constructed by considering the distribution of variances averaged over the phases, polarizations, inclinations, and sky locations
\bea\label{eq:sigma0}
    \sigma_k^2 
    &\equiv \langle |\tilde{\delta t}_{k}|^2 \rangle_{\delta, \psi, \imath, \hat{k}} \\
    &= \frac{1}{60 \pi^2} \sum_{j=1}^N \frac{A_j^2}{f_j^2} \left[(w_{k}^+)^2 + (w_{k}^-)^2 \right]\\
    &= \frac{\rho_c}{6\pi^3} \sum_{j=1}^N \frac{\Omega_j}{f_j^4} \left[ (w_{k}^+)^2 + (w_{k}^-)^2 \right] \,.
\eea
The last expression shows that this variance is given by an incoherent sum over sources, with each term directly associated with the energy emitted by the source $j$,
\be\label{eq:Omega_j}
    \Omega_j = \frac{\pi f_j^2}{10\rho_c} A_j^2 \,,
\ee
without including the possibility of destructive interference, as done in~\cite{Ellis:2023dgf}. We show the distribution of $\sigma_k^2$ in Fig.~\ref{fig:sigma0}. In the same way as above, and as in~\cite{Ellis:2023dgf}, the distribution is computed by dividing the sources into weak and strong ones and adding an analytic high-$\sigma_k$ tail.

Different realizations of SMBH binary masses and redshifts correspond to different $\sigma_k^2$, so their distribution $\td P/\td \sigma_k^2$ characterizes how the variances vary over the ensemble of SMBH binary populations. If the timing residuals are Gaussian due to variations in phases, polarizations, inclinations, and sky locations, the distribution of timing residuals is~\cite{Xue:2024qtx}\footnote{This approximation was referred to as Gaussian convolution in~\cite{Xue:2024qtx}.}
\be\label{eq:VA_Gaussian}
    \frac{\td P_{\rm GA}}{\td \ln |\tilde{\delta t_k}|} 
    = \int \td \sigma_k^2 \, \frac{\td P}{\td \sigma_k^2} \,\frac{|\tilde{\delta t_k}|^2}{\sigma_k^2} \exp\!\left[-\frac{|\tilde{\delta t_k}|^2}{2\sigma_k^2}\right]\,.
\ee
However, since the kurtosis is generally non-vanishing even when fixing the SMBH masses and redshifts and varying only phases, polarizations, inclinations, and sky locations~\cite{Lamb:2024gbh}, this estimate is not exact, as was already noted in~\cite{Xue:2024qtx}. As seen from Fig.~\ref{fig:histograms}, the variance-averaged Gaussian approximation~\eqref{eq:VA_Gaussian} provides a good estimate of the true distribution. In all cases we considered, we find a maximal deviation of about 20\% between the variance-averaged Gaussian~\eqref{eq:VA_Gaussian} and the true estimate. 

As a result, the distribution $\td P/\td \sigma_k^2$ can be used to approximate the inherent non-Gaussianity when performing statistical inference on SMBH models using PTA searches of Gaussian GW backgrounds (see e.g.~\cite{Ellis:2023dgf, Ellis:2023oxs, Raidal:2024odr, Sato-Polito:2024lew}). We will return to this point in Sec.~\ref{sec:AG_inference}.

\subsection{The \texttt{GWADpy} package}
\label{sec:package}

\begin{figure}
  \centering
  \includegraphics[width=\columnwidth]{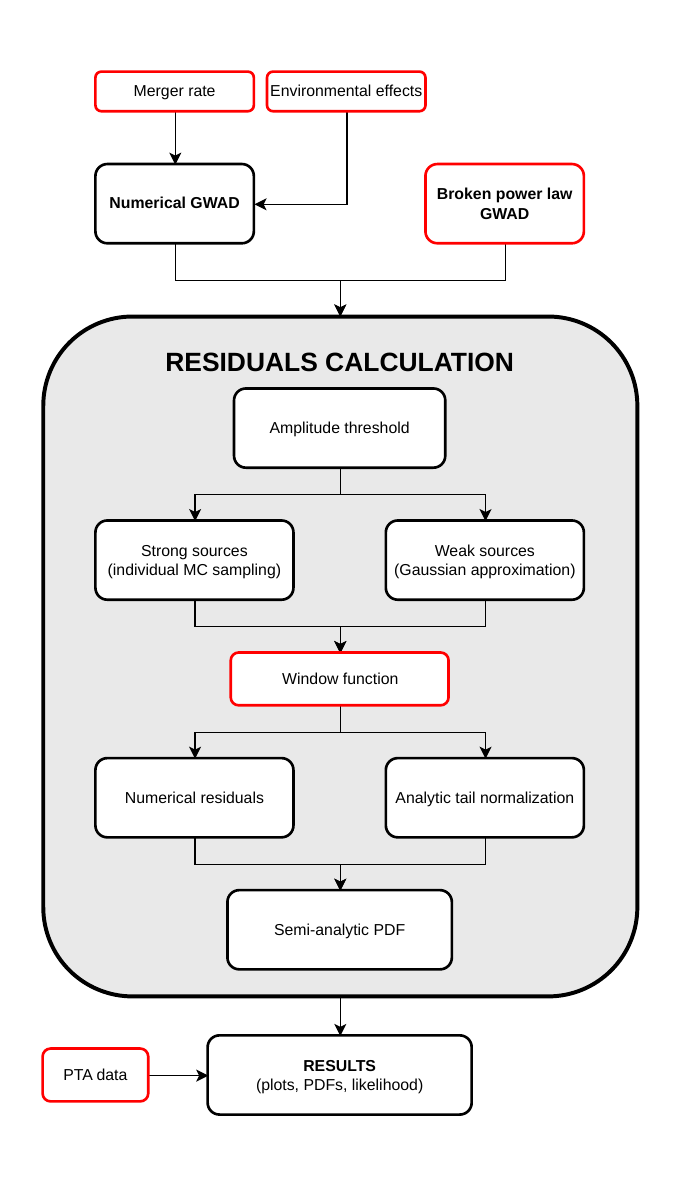}
  \caption{Schematic structure of the \texttt{GWADpy} code. The possible inputs are highlighted in red.}
  \label{fig:flowchart}
\end{figure}

To facilitate testing models of different SMBH binary populations, we have implemented the computational pipeline presented in this paper as a Python package, \texttt{GWADpy}~\cite{Raidal_GWAD_2026}. The structure of the code is illustrated in Fig.~\ref{fig:flowchart}. The input of the code is either parameters of the merger rate from Model I or Model II, from which a GWAD is computed, or a GWAD in the form of a broken power law
\be
     \frac{\td N}{\td A} 
     = \frac{N_b (p+q)^s}{\left[q \left(A/A_b\right)^{p/s}+p\left(A/A_b\right)^{q/s}\right]^s} \,,
\ee
where $N_b$ is the normalization, $p$ and $q$ are the asymptotic powers, $A_b$ is the power-law breaking point, and $s$ determines the smoothness of the break. As discussed in Sec.~\ref{sec:high-A}, $q$ should be fixed to 4.

\begin{figure}
  \centering
  \includegraphics[width=\linewidth]{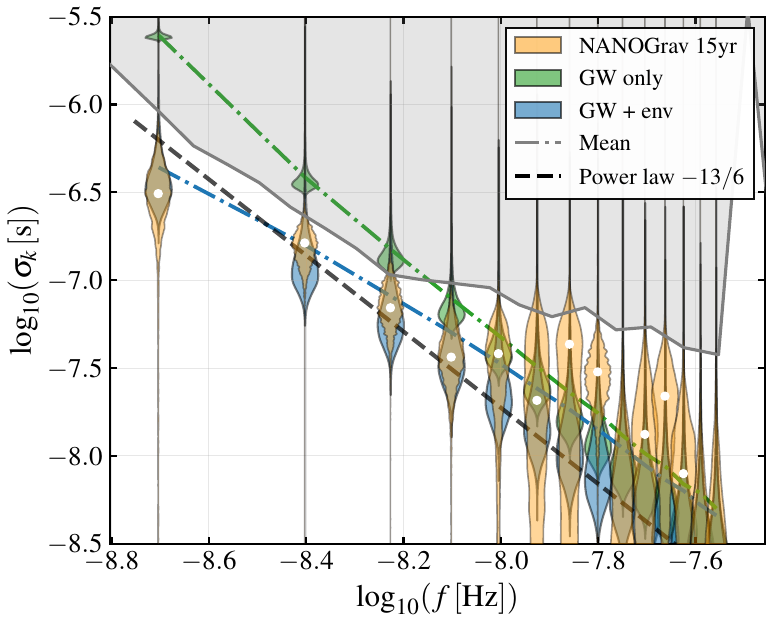}
  \caption{Violin plot showing the distributions of variances~\eqref{eq:sigma0} of timing residuals generated with \texttt{GWADpy} for purely GW driven binaries (green) and including also energy loss by the environmental effects (blue). The yellow violins show the posteriors of the NANOGrav 15-year data analysis. The model parameters are the same as in Fig.~\ref{fig:histograms}. The gray shaded region corresponds to the NANOGrav 15-year bound on single-source detection~\cite{NANOGrav:2023pdq}.}
  \label{fig:violins}
\end{figure}

Following the construction in Sec.~\ref{sec:stat_timing_residuals}, the GWAD is then used to divide the signal into strong and weak sources, and the timing residuals are calculated explicitly for strong sources and with a Gaussian approximation for weak sources. A convergent PDF for timing residuals can be obtained with $\mathcal{O}(10^5)$ realizations, as the high-amplitude tail is analytically attached to the PDF, which reduces the number of samples needed to resolve the total distribution. The package outputs the PDF of the timing residuals, which can be plotted or used to analyze the PTA data. The outputs are illustrated in Figs.~\ref{fig:histograms}, \ref{fig:sigma0}, and \ref{fig:violins}. There is also the option to output the variance-averaged likelihoods using the NANOGrav 15-year free-spectrum GW background posteriors~\cite{NANOGrav:2023gor}.

\section{Discussion}
\label{sec:discussion}

\subsection{Single loud source principle}
\label{sec:SLS}

Each frequency $f_k$ bin receives contributions from thousands of sources. Despite this, the GW signal of SMBHs is usually dominated by a small number of the loudest SMBH binaries. While this property is well-known in the literature, it is often generically attributed to the discrete nature of the sources or Poisson fluctuations~\cite{Sesana:2008mz,NANOGrav:2023hfp,Agazie:2024jbf,Xue:2024qtx,Sesana:2025udx}. However, these factors are not sufficient to explain why the SMBH signal is dominated by a handful of sources or the non-Gaussianity of the signal. The properties of the distribution of GWs from the SMBH binary population are characterized by the GWAD.

The GWAD has a heavy tail~\eqref{eq:A-4_tail}, which is also inherited by the amplitude timing residuals. Due to its power-law tail, it belongs to the class of subexponential distributions and is thus subject to the single big jump principle~\cite{Chistyakov1964} (see, e.g.~\cite{Foss2013} for a recent introduction). This principle states that when the sum of subexponential random variables $x_i$ exceeds some large number $X$, it is dominated by the maximal term in the sum.
\be\label{eq:single_big_jump}
    P\left( \textstyle\sum_i x_i > X \right) \sim P(\max \{x_i\} > X)\,,
    \quad
    X\to\infty \,.
\ee
In the context of GW backgrounds, we can rephrase it as the \emph{single loud source principle}, meaning that the loudest signals are more likely to be dominated by a few or even a single source. 

The single loud source principle is ultimately a property of the tail of the distribution of a large number of sources. It is the reason why the distribution of timing residuals from a population of binaries inherits the tail of the distribution~\eqref{eq:dt_tail} for a single binary.  In detail, there are $N$ ways to pick the maximal element in Eq.~\eqref{eq:single_big_jump}, so it holds that $P(\max \{x_i\} > X) \approx N P(x_i > X)$. This means that the tail of the probability distribution of the sum must be proportional to the number density of sources, which is exactly what we estimated in~\eqref{eq:dt_tail}.

Importantly, if the single loud source principle did not hold, it would be more likely that loud signals are composed of many weak ones with comparable strengths. For instance, if the GWAD exhibited a Gaussian tail, we would have that
\be
    P\left( \textstyle\sum^N_{i=1} x_i = X \right) \propto P(x_i = X/N)\,,
\ee
and the most likely configuration would be the one where all sources are equally strong. Although such scenarios are discrete and would display Poisson fluctuations, they would not lead to the domination of a few loud sources, indicating that discreteness is not a sufficient condition for the non-Gaussianity of the SMBH signal.

\subsection{Divergence of higher moments}

The power-law tail of the PDF of $|\tilde{\delta t}_k|$ implies that the statistical moments of order three and higher diverge. One way to regularize this divergence is to impose a cutoff in parameter space, restricting to SMBH binary populations with $A < A_{\rm cut}$. This induces a cutoff on the timing residuals, $|\tilde{\delta t}_k| < |\tilde{\delta t}_k|_{\rm cut}$, leading to
\be
    \langle |\tilde{\delta t}_k|^{n} \rangle \propto \begin{cases}
        |\tilde{\delta t}_k|_{\rm cut}^{n-3} \,, & n > 3 \\
        \ln |\tilde{\delta t}_k|_{\rm cut} \,, & n=3 
    \end{cases}\,.
\ee
Therefore, for $n \geq 3$, the regularized moments are dominated by the arbitrary cutoff scale and do not provide meaningful information about the underlying SMBH population.

This divergent behavior can be easily missed in the Poisson sampling of a binned SMBH parameter space, as implemented in the \texttt{holodeck} framework used in NANOGrav analyses~\cite{NANOGrav:2023hfp}. In such approaches, the merger rate is evaluated on a finite grid in redshift, which implicitly imposes a low-redshift cutoff. Consequently, the heavy tail~\eqref{eq:A-4_tail} can be suppressed by discretization artifacts rather than by the physics of the SMBH population, and estimates of higher moments (see, e.g.~\cite{Lamb:2024gbh,Lamb:2025niq}) risk systematically mischaracterizing the non-Gaussianity of the GW background and biasing inference on the SMBH merger rate. 

It is clear that empirical cutoffs on $A$ exist, as we do not observe heavy SMBH binaries in our local galactic neighborhood. The cutoffs implied by electromagnetic data are weak. For instance, the non-observation of SMBH binaries within the Local Group sets $D_L \gtrsim 2\,\mathrm{Mpc}$, while the mass of local SMBHs is bounded by $M \lesssim 2.7 \times 10^{11}\,\Msun$~\cite{2016MNRAS.456L.109K}. Together these give the weak upper bound $A \lesssim 10^{-9}\,(f/\mathrm{nHz})^{2/3}$ on the GW amplitude. Stronger constraints arise from PTA single source searches\footnote{These searches target coherent, deterministic signals from individual binaries rather than a stochastic superposition of unresolvable sources and are thus distinct from the GW background analysis.} They constrain $A \lesssim 10^{-14}$ in the $1-100$\,nHz range~\cite{NANOGrav:2023pdq}. However, as seen from Fig.~\ref{fig:violins}, these single source amplitude constraints exceed the total GW background amplitudes relevant for the PTA searches for GW backgrounds. Therefore, while such cutoffs can formally regulate ensemble averages, they would not significantly affect the statistical inference of the GW background.

Moreover, if the most stringent constraint on $A$ is obtained from the same PTA dataset that is used to limit the GW background, then the resulting bounds on $A$ for the SMBH population and for the GW background may not be statistically independent. One way to address this is to perform a joint analysis of individually resolvable sources and the GW background, so that the upper limit on $A$ for the SMBH population corresponds directly to the upper limit on the amplitude of the resolvable sources.

\subsection{Variance within and across realizations}

It is important to distinguish between the fluctuations of timing residuals within a specific realization of the SMBH binary population and the variability of timing residuals across the full ensemble of possible realizations. Higher-order moments of $\delta t_k$ estimated from a single realization are necessarily finite and are thus not represented by the ensemble averages. Moreover, despite having a finite ensemble average, the variance of timing residuals $\langle|\tilde{\delta t}_k|^2\rangle$ (or equivalently, the GW energy density spectrum $\Omega_{\rm GW}(f_k)$) can exhibit large realization-to-realization fluctuations because the size of these fluctuations is controlled by $\langle |\tilde{\delta t}_k|^4\rangle$, which diverges.

The variability of the signal across realizations characterizes our lack of information about the sources. Thus, even if it happens to be the case that the heavy tail may not be directly observable within a single realization, it will nonetheless impact statistical inference on SMBH models. A Gaussian likelihood does not account for the possibility that the signal is dominated by a few loud binaries and will therefore underestimate the probability of such configurations. This suggests that the appropriate treatment is to work with the full distribution of timing residuals rather than to characterize the background through its low-order moments alone.

Let us briefly examine the extent to which the variability of the ensemble might be observable from the realization of the SMBH population in our Universe. In this paper, we focus on the distribution for a single pulsar. When considering how the timing residuals vary from pulsar to pulsar, the only change is in the relative sky location between the pulsar and the source. However, with multiple pulsars, we will also have access to correlations between timing residuals, which is crucial to determining the GW origin of the signal through the Hellings-Downs curve~\cite{Hellings:1983fr}, but can also improve our access to the variability in phases, inclinations, and polarizations encoded in the response~\eqref{eq:response}. 

Clearly, increasing the number of pulsars does not affect the realization of GW amplitudes in our Universe, which combines all available information about SMBH masses and redshifts. In the ideal case, where we are able to resolve a sufficiently large set of the loudest sources, the GWAD can be partially resolved for smaller amplitudes at which the signal is expected to be composed of multiple sources. However, our access to the shape of the heavy tail will always be limited by cosmic variance. Specifically, the region above which less than a single event is expected can never be fully sampled directly within a single realization of the Universe due to the large Poisson fluctuations in the expected number of sources. However, the universal $A^{-4}$ shape~\eqref{eq:A-4_tail} can allow for its reconstruction by extrapolation.

\subsection{SMBH model inference}
\label{sec:AG_inference}

Current PTA analyzes model the GW-induced timing residuals as a Gaussian stochastic process and infer posteriors for the variance at each Fourier mode (in a free-spectrum fit) or under the assumption of a power-law spectral shape. In SMBH model inference, the NANOGrav collaboration uses the \texttt{holodeck} framework to generate large ensembles of realizations, from which the mean and variance of the incoherent sum of the signals, $\sum_j A_j^2$, are estimated in each frequency bin and used to construct a log-normal likelihood that is compared to the free-spectrum posteriors~\cite{NANOGrav:2023hfp}. By contrast, the EPTA collaboration performs the inference using the mean spectrum predicted for each choice of population parameters, neglecting the realization-to-realization scatter, as they explicitly state in Ref.~\cite{EPTA:2023xxk}. This scatter is included in a separate qualitative check, in which the spectrum of each simulated realization is fit with a power law and the resulting distribution is compared with the data in the amplitude-spectral index plane.

An improved approach was developed in~\cite{Ellis:2023dgf} using the full non-Gaussian distribution of the energy density spectrum  
\be
    \Omega_{\rm GW}(f_k) = \frac{1}{\ln(f_{k+1}/f_k)} \sum_{j=1}^{N(f_k)} \Omega_j \,,
\ee
where $N(f_k)$ denotes the number of binaries in the $k$th Fourier bin. The likelihood is constructed by combining this distribution with the free-spectrum posteriors $p_k(\Omega_{\rm GW})$ obtained from the Gaussian process PTA analysis:
\be \label{eq:likelihood}
    \mathcal{L}(\vec{\theta}) = \prod_k \int \td \Omega_{\rm GW}  \frac{\td P(\Omega_{\rm GW}|f_k,\vec{\theta})}{\td \Omega_{\rm GW}} \, p_k(\Omega_{\rm GW}) \,.
\ee
In the NANOGrav analysis~\cite{NANOGrav:2023hfp}, ${\td P(\Omega_{\rm GW})/\td \Omega_{\rm GW}}$ is effectively approximated by a log-normal distribution, while the EPTA analysis~\cite{EPTA:2023xxk} assumes a delta distribution and therefore ignores the spread. In contrast, the method of~\cite{Ellis:2023dgf} preserves the complete non-Gaussian structure.

For a top-hat window function, commonly assumed in SMBH model inference~\cite{NANOGrav:2023hfp,EPTA:2023xxk,Ellis:2023dgf}, the energy density spectrum is proportional to the variance of the timing residuals averaged over the phases, polarizations, inclinations, and sky locations, as given in Eq.~\eqref{eq:sigma0}, $\Omega_{\rm GW}(f_k) \propto \sigma_k^2$. As shown in Sec.~\ref{sec:AG}, the variance-averaged Gaussian, obtained by marginalizing over the realization-to-realization distribution of $\sigma_k^2$, provides a good approximation to the full timing residual distribution. This means that, for a fixed realization, i.e., fixed $\sigma_k^2$, the timing residuals are approximately Gaussian, which is precisely the assumption underlying the PTA Gaussian-process analysis and its free-spectrum posteriors $p_k(\Omega_{\rm GW})$. As pointed out in~\cite{Xue:2024qtx}, this validates the factored structure of the likelihood~\eqref{eq:likelihood}: the Gaussian-process PTA posteriors correctly describe the data at fixed $\Omega_{\rm GW}$, and can therefore be consistently combined with the non-Gaussian population prior ${\rm d}P/{\rm d}\Omega_{\rm GW}$. An example of this procedure is illustrated in Fig.~\ref{fig:violins} (using $\sigma_k^2$  instead of $\Omega_{\rm GW}$), where the data (yellow violins) should be compared with the theoretical estimates (blue and green violins) from specific SMBH population models. We note that while this approximation is well justified for the single-pulsar likelihood, further work is required to assess its validity for inter-pulsar correlations.

Importantly, as discussed above, the non-Gaussian heavy tail is a property of the ensemble of SMBH populations and quantifies our degree of uncertainty. Since we observe a single realization of the Universe, this tail is not directly measurable, especially in the region where we expect fewer than one source. Despite that, including it is relevant for unbiased SMBH model inference. Although approximate, the variance-averaged likelihood~\eqref{eq:likelihood} provides a simple way to incorporate it into existing Gaussian analyzes.

\section{Conclusions}
\label{sec:concl}

We have investigated the non-Gaussian properties of the gravitational wave (GW) background generated by a population of inspiraling supermassive black hole (SMBH) binaries. We have shown that the GW amplitude distribution (GWAD) exhibits a universal broken power-law structure: a heavy high-amplitude tail scaling as $\propto A^{-4}$, arising from the possibility of nearby sources, and a low-amplitude regime that encodes the SMBH merger rate and the underlying energy-loss mechanisms of the binaries. 

We have demonstrated that this heavy-tailed behavior propagates directly to the distribution of pulsar timing residuals. In particular, the timing residuals inherit the $\propto |\tilde{\delta t}_k|^{-4}$ tail, implying that statistical moments of order three and higher formally diverge. This highlights a fundamental limitation of characterizing the GW background using low-order moments or Gaussian statistics, as is often done in pulsar timing array (PTA) analyzes.

Our results establish that the nHz GW background from SMBH binaries is intrinsically non-Gaussian and governed by the single loud source principle, whereby individual Fourier modes are often dominated by a small number of loud sources, even though each mode receives contributions from thousands of binaries. Consequently, summary statistics such as the variance or kurtosis are not robust descriptors of the signal, as they are sensitive to rare realizations and implicit population cutoffs. A consistent statistical description, instead, requires modeling the full distribution of timing residuals, or equivalently, the underlying GWAD.

At the same time, we find that the variance-averaged Gaussian approximation provides an accurate description of the timing-residual statistics. This result justifies a factored likelihood approach, in which standard Gaussian-process PTA posteriors are combined with a non-Gaussian population prior derived from the GWAD. Such a construction enables a consistent incorporation of non-Gaussian effects into SMBH population inference without abandoning existing PTA analysis pipelines.

To facilitate such analyzes, we have developed a fast and flexible numerical framework implemented in the \texttt{GWADpy} package, which allows one to compute timing residual distributions directly from a given SMBH merger rate or GWAD~\cite{Raidal_GWAD_2026}. The framework incorporates both strong and weak source contributions, includes interference effects, and accounts for realistic window functions that determine the mapping between sources and Fourier modes. We have shown that data processing choices, such as whitening and filtering, play a crucial role in shaping inter-mode correlations and must be consistently incorporated into theoretical modeling.

Our findings have direct implications for PTA data analysis. Standard Gaussian likelihoods may fail to capture the true statistical properties of GWs from SMBH populations, potentially biasing the inference of the GW background and the underlying SMBH population. This motivates moving beyond Gaussian assumptions and incorporating non-Gaussian statistics at the likelihood level, for example, through GWAD-based approaches.

\begin{acknowledgments}
We thank G.~Franciolini, J.~El~Gammal and M.~Pieroni for insightful discussions. This work was supported by the Estonian Research Council grants PSG869, TARISTU24-TK3, TARISTU24-TK10, and the Centre of Excellence programme TK202 of the Estonian Ministry of Education and Research. The work of V.V. was partially funded by the European Union's Horizon Europe research and innovation program under the Marie Sk\l{}odowska-Curie grant agreement No. 101065736.
\end{acknowledgments}

\appendix

\section{Response function} 
\label{app:response}

\subsection{Distribution of the response} 

Here we show that the quantity $e^{i\delta_j}R_{J,j}$ in Eq.~\eqref{eq:deltat} can be characterized by two independent random variables: the overall phase $\bar \delta_{J,j} \equiv \delta_j + \arg R_{J,j}$ and the modulus $|R_{J,j}|$. Since we are working with a single source and one pulsar, we drop the indices labeling the sources and pulsars below. 

First, the independence of the total phase is easily checked by noting that, since $\delta$ is uniformly distributed, $p(\delta) = 1/(2\pi)$, the distribution of the overall phase $\bar \delta$ conditioned on $|R|$ is also uniform,
\be
    p(\bar \delta) = \int \td \arg R \, p(\bar \delta - \arg R) p(\arg R|\,|R|) = \frac{1}{2\pi}\,,
\ee
and thus independent of $|R|$.

Second, choosing the coordinates so that $\vec{u} = (0,0,1)$ and $\hat{k} = (\sin \theta \cos \phi, \sin \theta \sin \phi, \cos\theta)$ reduce the antenna pattern functions to $F^{+} = \sin^2(\theta/2) \cos(2\psi)$ and $F^{\times} = \sin^2(\theta/2) \sin(2\psi)$. Consequently, we get
\bea \label{eq:absR2}
    |R|^2 
    &= \frac{1}{4} \left[1\!-\!\cos(2\pi f L (1\!+\!\cos\theta)) \right] \sin^4(\theta/2) \\
    &\quad\times \left[ 1\!+\!6\cos^2 \imath\!+\!\cos^4 \imath\!+\!\sin^4 \imath \cos(4\psi) \right] \,,
\eea
which implies that $|R| \leq 2$. Averages over source sky locations and polarizations yield $\langle F^\lambda \rangle = \langle R \rangle_{\psi, \hat{k}} = 0$ and $\langle F^\lambda F^{\lambda'} \rangle = \delta^{\lambda\lambda'}/6$, while averaging also over binary inclinations results 
\be\label{eq:R2_mean}
    \langle |R|^2 \rangle = \frac{4}{15} + \frac{{\rm sinc}(4\pi f L)-1}{10(\pi f L)^2} \,.
\ee
The last term suppresses the response when $f L \lesssim 0.5$. For all pulsars $L > 100 \,{\rm pc}$, so the frequency dependence of $|R|$ can be safely neglected for $f \gtrsim 0.1\,{\rm nHz}$. As shown in Appendix~\ref{app:window_f}, signals at sub-nHz frequencies are strongly suppressed by the window function. We therefore adopt the distribution of $|R|$ in the limit $f L \gg 1$, as derived below.

To estimate the distribution of $|R|$, we first recast Eq.~\eqref{eq:absR2} as
\be
    |R|^2 = |R_0(\cos\theta, f L)|^2 T(\imath,\psi)^2\,,
\ee
where
\bea
    |R_0|^2 
    &\equiv [1 - \cos(\pi K (1+y))]\frac{(1-y)^2}{2}\,,     
    \\
    |T|^2 
    &\equiv \frac{1}{8}\left[ 1\!+\!6z^2 \!+\!z^4 \!+\!(1-z^2)^2 \cos(4\psi) \right]\,,
\eea
that factors $|R|$ into a contribution at zero inclination $\imath = 0$ and the inclination dependent part, and we defined $K \equiv 2 f L$ and the variables $y \equiv \cos\theta$, $z \equiv \cos \imath$. The variables are uniformly distributed, and given the symmetries of $|R|$, it is sufficient to consider them in the ranges $y \in [-1,1]$, $z \in [0,1]$, $\psi \in [0,\pi/4]$.

\begin{figure}
    \centering
    \includegraphics[width=0.95\columnwidth]{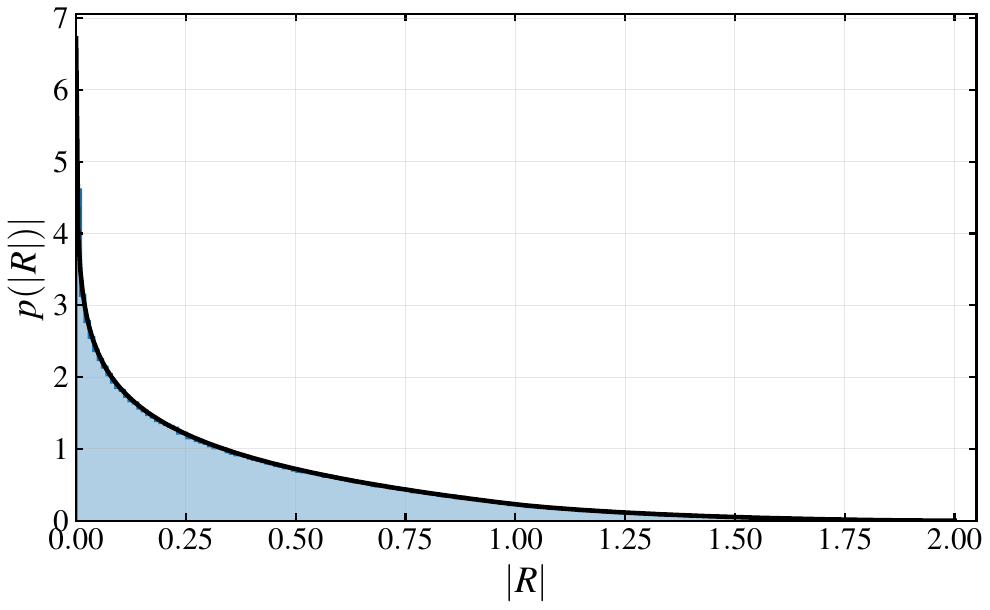} 
    \caption{Distribution of $|R|$. The histogram shows numerical samples while the black line shows~\eqref{eq:pR}.}
    \label{fig:R_PDF}
\end{figure}

We estimate the distribution of $|R_0|$ in the limit $K \gg 1$. Formally, it is given by
\be
    p(|R_0|^2) = \int^{1}_{-1} \frac{\td y}{2} \delta (|R_0|^2 - |R_0|^2(y))\,.
\ee
So, we need to invert the mapping $y \mapsto |R_0|$ by considering intervals of $|R_0|$ where it is one-to-one. In the $K \gg 1$ limit, these intervals correspond to $k/K \leq 1+y < (k+1)/K$, where $k$ is an integer in the range $0 \leq k < 2K$. For simplicity, we can further assume that $K$ is an integer, as the contribution from the fractional part of $K$ tends to 0 in the $K \to \infty$ limit. In each of the narrow intervals, it is sufficient to consider only the variation of the cosine; thus, the distribution is
\be
    p(|R_0|^2|k) = \frac{ \Theta \left( (1-y_k)^2 - |R_0|^2 \right)}{\pi \sqrt{\left((1-y_k)^2 - |R_0|^2\right)|R_0|^2}}\,,
\ee
where $\Theta$ is the step function and $y_k \approx k/K-1$. The probability of being in an interval $k$ is proportional to its width, i.e., $p(k) \equiv p(k/K \leq 1+y < (k+1)/K) = 1/(2K)$. Therefore,
\bea
    p(|R_0|^2) 
    &= \sum^{2K-1}_{k=0} p(|R_0|^2|k) p(k)
    \\
    &\to \int^{1}_{-1} \frac{\td y}{2} \frac{ \Theta \left((1-y)^2 - |R_0|^2 \right)}{\pi \sqrt{\left((1-y)^2 - |R_0|^2\right)|R_0|^2}}
    \\
    &= \frac{1}{2\pi \sqrt{|R_0|^2}}{\rm arccosh}\left(\frac{2}{\sqrt{|R_0|^2}}\right)\,,
\eea
where the arrow indicates the $K\to \infty$ limit in which the sum can be approximated by an integral. Importantly, the dependence on $K$ and thus the source frequency $f_j$ vanishes in that limit. The distribution of $|R_0|$ is therefore
\be
    p(|R_0|) = \frac{1}{\pi }{\rm arccosh}\left(\frac{2}{|R_0|}\right)\,,
\ee
with $|R_0| \in [0,2]$. 

Finally, the distribution of $|R|$ can be given by a double integral
\bea\label{eq:pR}
    p(|R|) 
    &= \int^{1}_{0} \td z \int^{\pi/4}_{0} \frac{\td \psi}{\pi/4} \int^{2}_{0} \td |R_0| \, p(|R_0|) 
    \\
    &\qquad\quad\times \delta\big(|R| - |R_0| |T|(z,\psi)\big)
    \\
    &= \frac{4}{\pi^2}\int^{1}_{0} \td z \int^{\pi/4}_{0} \td \psi \, \Theta\big(2|T|(z,\psi) - |R|\big)
    \\
    &\qquad\quad\times \frac{1}{|T|(z,\psi)}{\rm arccosh}\left(\frac{2 |T|(z,\psi)}{|R|}\right) \,.
\eea
Fig.~\ref{fig:R_PDF} shows this distribution. In our analysis, we evaluate it numerically.

Although the integral~\eqref{eq:pR} does not admit a closed-form expression, we find that $p(|R|)$ can be well approximated by $|\ln(|R|/2)|(2-|R|)/3$, which may be useful in contexts where rapid or analytic evaluation is needed. The approximation introduces inaccuracies on the order of a few percent and, in particular, it yields $\langle|R|^2\rangle = 7/27$, which is 3\% lower than the exact value $4/15$.

\subsection{Response of quasi-monochromatic signals}

Consider now a quasi-monochromatic signal with a slowly evolving frequency. The polarization modes~\eqref{eq:pol_modes} are then recast as
\bea
	&h_j^+(t) = \frac{1+\cos^2\imath_j}{2} \, A_j(t) \cos(\Phi_j(t))\,, \\
	&h_j^\times(t) = \cos\imath_j \, A_j(t) \sin(\Phi_j(t))\,,
\eea
where $A_j(t)$ and $\Phi_j(t)$ denote the time-dependent amplitude and phase of the signal, respectively.  The derivative of the phase gives the instantaneous signal frequency $f_j \equiv \dot\Phi_j/(2\pi)$. The monochromatic case corresponds to a constant amplitude $A_j$ and frequency $f_j$, so that $\Phi_j(t) = 2\pi f_j t + \delta_j$.

By repeating the derivation in Sec.~\ref{sec:timing_residuals}, we find that the timing residual of pulsar $J$ is
\begin{align}
\label{eq:deltatJ_quasi-mc}
    \delta t_J 
    &= \sum^{N}_{j=1}\frac{\tilde R_{J,j}}{2} \int^t_{t_p} \td \tilde t \,  A_j(\tilde t) e^{i \Phi_j(\tilde t)} + {\rm c.c.} \,
    \\
    &\approx \sum^{N}_{j=1}\frac{\tilde R_{J,j}}{4\pi i} \left[ \frac{A_j(t)}{f_j(t)} e^{i \Phi_j(t)} - \frac{A_j(t_p)}{f_j(t_p)} e^{i \Phi_j(t_p)} \right] + {\rm c.c.} \,,\nonumber
\end{align}
where we redefined the variable of integration $\tilde t = t - (L_J-s)(1+\hat{u}_J\cdot\hat{k}_j)$, defined the pulsar time $t_p = t - L_J(1+\hat{u}_J\cdot\hat{k}_j)$, and introduced the response
\be
    \tilde R_{J,j} = \frac{1\!+\!\cos^2 \imath_j}{2} F_J^+(\hat{k}_j,\psi_j)\!-\!i \cos \imath_j F_J^\times(\hat{k}_j,\psi_j) \,,
\ee
which does not include the rapidly oscillating prefactor of the monochromatic response function~\eqref{eq:response}. Instead, this prefactor is now expressed explicitly in Eq.~\eqref{eq:deltatJ_quasi-mc}.

To better probe the boundaries of the monochromatic assumption, let us explicitly derive the second line in Eq.~\eqref{eq:deltatJ_quasi-mc} under the assumption that $f_j$ and $A_j$ vary slowly in time. The integral over the path of the pulse gives
\begin{widetext}
\bea\label{eq:int_quasi-mc}
       \int^t_{t_p} \td \tilde t \, A_j(\tilde t) e^{i \Phi_j(\tilde t)}
    &= \int^t_{t_p} \td \tilde t \, \frac{A_j(\tilde t)}{i \Phi'_j(\tilde t)} \,\pd_{\tilde t} e^{i \Phi_j(\tilde t)}
    = \frac{A_j(\tilde t)}{i \Phi'_j(\tilde t)} e^{i \Phi_j(\tilde t)} \bigg|^t_{t_p} 
    - \int^t_{t_p} \td \tilde t \,  \pd_{\tilde t}\!\left(\frac{A_j(\tilde t)}{i \Phi'_j(\tilde t)}\right) e^{i \Phi_j(\tilde t)}
    \\
    &= \frac{A_j}{2\pi i f_j} e^{i \Phi_j(t)} \left[1 - \left(\frac{f_j(t)}{f_{j}(t_p)}\right)^{\!1/3}e^{-i (\Phi_j(t) - \Phi_j(t_p))} \right] + \mathcal{O}(\dot f_j/f_j^2)\,,
\eea
\end{widetext}
where, in the last line, we used $A_j \propto f_j^{2/3}$ and indicated that the remaining integral contributes only a subleading correction. In the monochromatic limit, the prefactor in front of the square brackets reproduces the harmonic oscillations in~\eqref{eq:deltatJ}, while the bracketed term gives the oscillatory term in the response~\eqref{eq:response} for monochromatic signals. Overall, Eq.~\eqref{eq:int_quasi-mc} gives the standard result that the timing residual produced by an inspiralling binary can be written as the difference between the Earth and pulsar terms~\cite{Estabrook:1975jtn, Wahlquist:1987rx}. We remark that the boundary term could be refined further by iterating the partial integration.

We find that the monochromatic limit works well when the following conditions are met: First, in order to neglect the integral in the second row of Eq.~\eqref{eq:int_quasi-mc}, the frequency and amplitude evolution should be adiabatic 
\be\label{eq:cond_ad}
    \dot f_j/f_j^2 \ll 1\,.
\ee
This implies also that the variation of the amplitude is adiabatic $\dot A_j/(A_j f_j) = \dot f_j/(3f_j^2) \ll 1$ since $A_j \propto f_j^{2/3}$.

Second, in order for the pulsar term not to differ much from the monochromatic case we would need that $f_{j}(t_p) \approx f_j(t)$. Specifically, the phase difference 
\begin{align}
    &\Phi_j(t) - \Phi_j(t_p)
    \\
    &= {\rm const.} + L_{J}(1+\hat{u}_J\cdot\hat{k}_j)\dot f_j(t_0) (t-t_0) + \ldots\nonumber
\end{align}
should not vary significantly during the observational period. Here $t_0$ denotes a reference time around which the phase is expanded, e.g., the beginning of observations. This condition on the phase implies that 
\be\label{eq:cond_cond}
    \dot f_j L_{J} T \ll 1\,
\ee
or, equivalently, that the pulsar term should not drift into a neighboring frequency bin, i.e., $L_{J}\dot f_j \approx  f_j(t) -  f_j(t_p) \ll 1/T$. Since $L_{J} T f_j^2 \gg 1$, the condition on the phase \eqref{eq:cond_cond} is strictly stronger than the adiabaticity condition~\eqref{eq:cond_ad}. In Sec.~\ref{sec:timing_residuals} we showed that $\dot f_j L_{J} T \approx \mathcal{O}(0.1)$ for typical sources.

Let us briefly consider the implications for the distribution \eqref{eq:pR} of $|R|$. We checked numerically that the mean $\langle |R|^2 \rangle$ agrees well with its expected value 4/15 in the $L_{J} f_j\gg 1$ limit as long as $L_{J}\dot f_j/f_j \lesssim 0.5$. In this case, the pulsar term does not contribute to $\langle |R|^2 \rangle$, but it will still affect the distribution as it does in the monochromatic case. 

Finally, the random nature of $L_{J}(1+\hat{u}_J\cdot\hat{k}_j)$ in the pulsar term completely dominates the variability of the phase. Therefore, it reduces the relevance of small frequency changes during the observational period in searches of stochastic GW backgrounds. Moreover, it has been argued that it can be dropped also in searches for single sources because the phases vary from pulsar to pulsar causing them to cancel if the number of pulsars is sufficiently large~\cite{Ellis:2012zv}.

\section{Window functions}
\label{app:window_f}

\begin{figure}[t]
  \centering
   \includegraphics[width=\columnwidth]{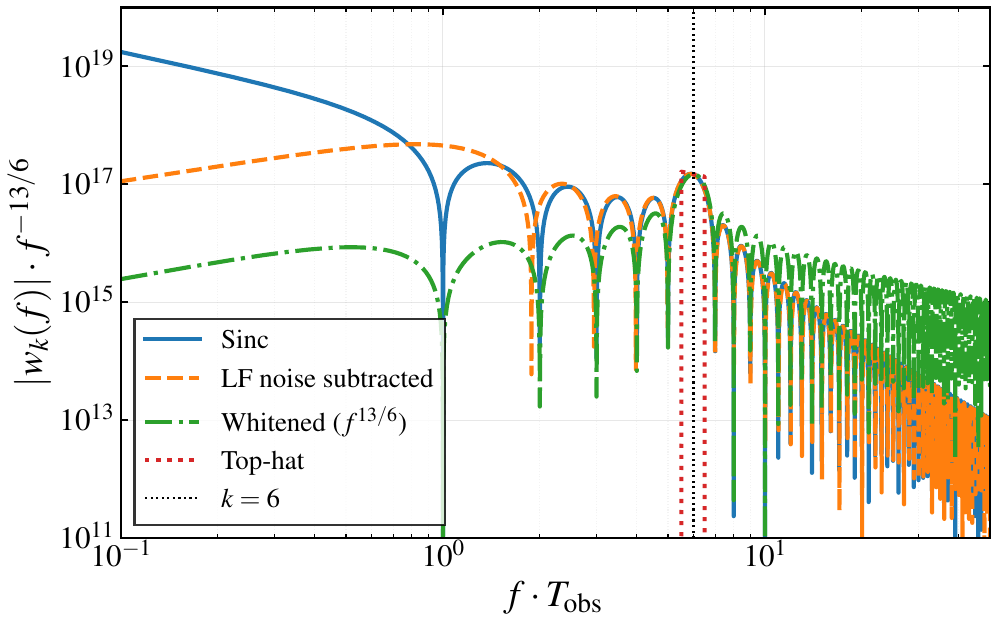}
  \caption{Window function for the mode $k = 6$ multiplied by $f^{-13/6}$, which corresponds to the scaling of the standard deviation of the timing residual from SMBH binary populations.}
  \label{fig:window}
\end{figure}

Direct computation of the Fourier coefficients of the timing residuals~\eqref{eq:deltatJ} gives
\bea
    \tilde{\delta t}_k 
    &= \frac{1}{T} \int_{-T/2}^{T/2} \td t \, \delta t(t) e^{-2\pi i f_k t} \\
    &= \sum_j \left[ X_j w_k(f_j) + X_j^* w_k(-f_j) \right] \,.
\eea
where 
\be
    X_j = \frac{A_j R_{J,j}}{4\pi i f_j} \, e^{i \delta_j}
\ee
and the window function is
\be
    w_k(f) = \sinc[\pi T (f-f_k)] \,.
\ee
The red curve in Fig.~\ref{fig:window} shows the sinc window function for $k=6$, multiplied by the scaling of $\sqrt{\langle |X_j|^2\rangle} \propto f^{-13/6}$. This indicates that sources with $Tf < 1$ can provide the dominant contribution to all Fourier modes. In the following, we discuss how this window function is modified through data processing. 

\subsection{Low-frequency noise subtraction}

Deterministic slowly changing components cannot generally be distinguished from low-frequency (LF) background noise. Such terms are removed in the analysis of PTA data~\cite{Taylor:2021yjx}. We consider terms with at most quadratic dependence in time, so that
\be
    \delta t(t) = \delta t_0(t) - \sum^{2}_{n=0} A_n \left(\frac{t}{T}\right)^n \,,
\ee
where $\delta t_0(t)$ denotes the timing residual induced by GWs (see Eq.~\eqref{eq:deltatJ}). We choose the coefficients $A_n$ such that $\int^{T/2}_{-T/2} \td t \, \delta t(t)^2$ is minimized.\footnote{We assume uniform timing noise across the observation span, such that minimizing $\int \td t\,\delta t^2(t)$ corresponds to the standard unweighted inner product. Irregular cadence and heteroscedastic timing uncertainties instead lead to a noise-weighted inner product~\cite{vanHaasteren:2012hj}. The simplified treatment adopted here is intended solely for analytic tractability in the idealized window-function comparison and should not be interpreted as a realistic model of PTA noise properties.} Varying with respect to $A_n$ gives
\be
    \tau_0 = A_0 + \frac{A_2}{12}\,,\quad
    \tau_1 = \frac{A_1}{12}\,,\quad
    \tau_2 = \frac{A_0}{12}+ \frac{A_2}{80}\,,
\ee
where $\tau_n \equiv \int^{T/2}_{-T/2} \td t \, \delta t_0(t) t^n/T^{n+1}$. Solving for $A_n$ we find that
\bea
    \delta t(t) = \delta t_0(t) - \int^{T/2}_{-T/2} \frac{\td t'}{T} \, \delta t_0(t')K(t,t') \,,
\eea
where
\be
    K(t,t') = \frac{9}{4} + 12\frac{t t'}{T^2} - 15 \frac{t^2 + {t'}^2}{T^2} + 180 \frac{t^2{t'}^2}{T^4} \,.
\ee
The resulting window function for the mode $k$ is
\begin{widetext}
\bea \label{eq:w_substr}
    w_k(f) &= \sinc[\pi T(f-f_k)] - \int^{T/2}_{-T/2} \frac{\td t' \td t}{T^2} \,K(t,t') \, e^{-2\pi i (f_k t'-f t)} \\
    &= \sinc[\pi T(f-f_k)] +  \sin[\pi T(f-f_k)] \left[\frac{3}{(\pi T)^3 f^2 f_k} - \frac{15}{(\pi T)^3 f f_k^2} + \frac{45}{(\pi T)^5 f^3 f_k^2}
    \right] \\
    &\quad - \cos[\pi T(f-f_k)] \left[
    \frac{3}{(\pi T)^2 f f_k} + \frac{45}{(\pi T)^4 f^2 f_k^2}
    \right] \,.
\eea
\end{widetext}

The green curve in Fig.~\ref{fig:window} shows the window function~\eqref{eq:w_substr}, multiplied by the scaling $\sqrt{\langle |X_j|^2\rangle} \propto f^{-13/6}$. We see that LF noise subtraction suppresses the contribution from low-frequency binaries, but this suppression is insufficient to efficiently mitigate spectral leakage for spectra as steep as those expected from a population of SMBH binaries

\subsection{Whitening}

To mitigate spectral leakage, the time series can first be whitened, i.e., transformed so that the residual noise becomes approximately uncorrelated with unit variance.  A whitened time series $\delta t_W(t)$ is obtained by convolving the data $\delta t(t)$ with a filter kernel $W(t)$:
\be
    \delta t_W(t) = \int \td t' \delta t(t-t') W(t') \,.
\ee
The Fourier coefficients computed in the whitened domain are then mapped back to the original noise properties by post-coloring, which is obtained by dividing them by the Fourier transform of the whitening kernel evaluated at the corresponding Fourier mode frequency:
\be
    \tilde{\delta t}_{k} = \frac{1}{\tilde{W}(f_k)} \frac{1}{T} \int_{-T/2}^{T/2} \td t \,\delta t_W(t) e^{-2\pi i f_k t} \,.
\ee
This gives
\be \label{eq:w_white}
    w_k(f) = \frac{\tilde{W}(f)}{{\tilde{W}(f_k)}} \sinc[\pi T(f-f_k)] \,.
\ee

We note that the filter kernel $W(t)$ used to whiten the timing series is not known a priori. In practice, simple approximations such as first or second differences of the timing residuals can be used as discrete whitening filters without requiring an explicit estimate of the spectrum~\cite{Coles:2011zs}. In the idealized case, where the power spectral density $S(f)$ is exactly known, the whitening filter in the frequency domain is given by $\tilde{W}(f) = 1/\sqrt{S(f)}$. Nevertheless, the filtered series is not perfectly uncorrelated because finite sampling and the limited observation window introduce residual correlations with a characteristic sinc shape.

The blue curve in Fig.~\ref{fig:window} shows the window function~\eqref{eq:w_white} for $\tilde{W}(f) \propto f^{13/6}$, multiplied by the scaling $\sqrt{\langle |X_j|^2\rangle} \propto f^{-13/6}$. We see that, in this case, processing the timing series with pre-whitening and post-coloring efficiently suppresses leakage from low-frequency sources.

\begin{figure*}
    \centering
    \includegraphics[width=0.8\linewidth]{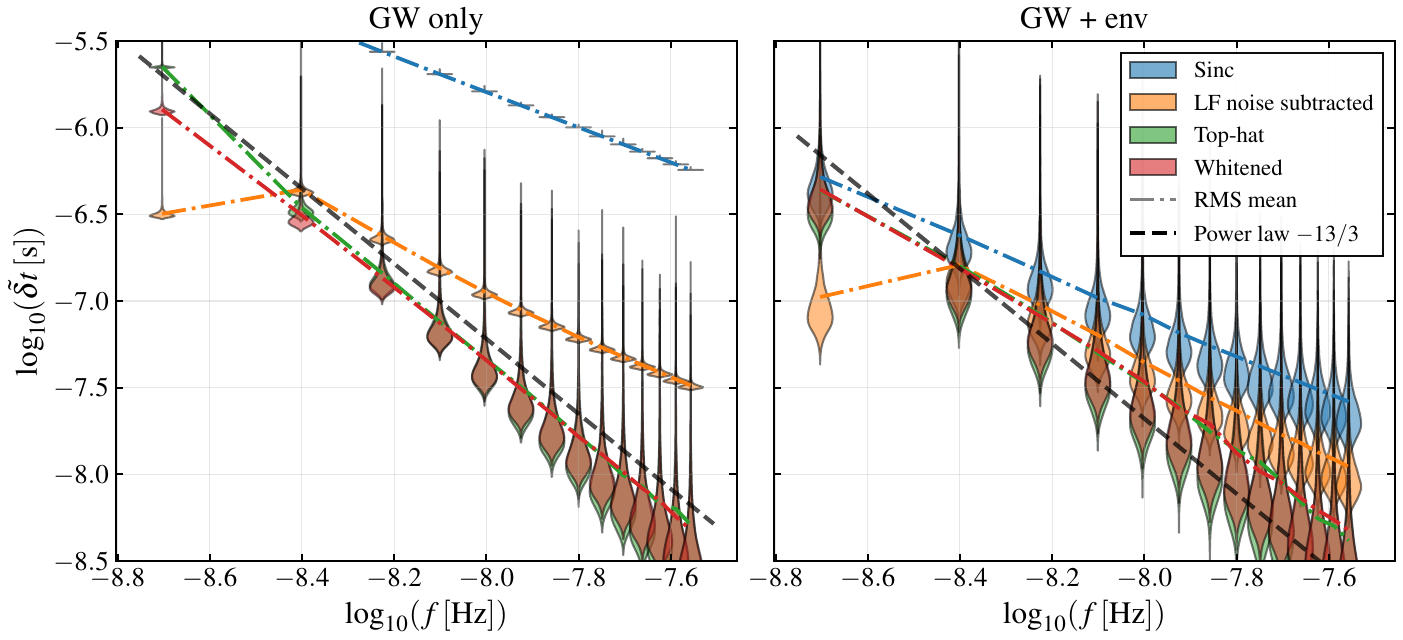}
    \caption{Violin plot showing the distributions of timing residuals obtained by using different window functions. The distributions are obtained from $10^6$ realizations, with $N_{\rm S}=50$. The source population is derived using Model I merger rate with $\{p_{\rm BH},a,b,\sigma\}=\{0.6,8.95,1.4,0.47\}$ and environmental effects (in the right panel) with $\{\alpha,\beta,f_{\rm ref}\}=\{8/3,5/8,30\,\rm{nHz}\}$. Note that we have placed a cutoff for low-frequency sources at $f=0.1 \rm{nHz}$. Lowering this would lead to an even higher amplitude signal in the sinc window case.}
    \label{fig:violins_windows}
\end{figure*}

\begin{figure*}
    \centering
    \includegraphics[width=\linewidth]{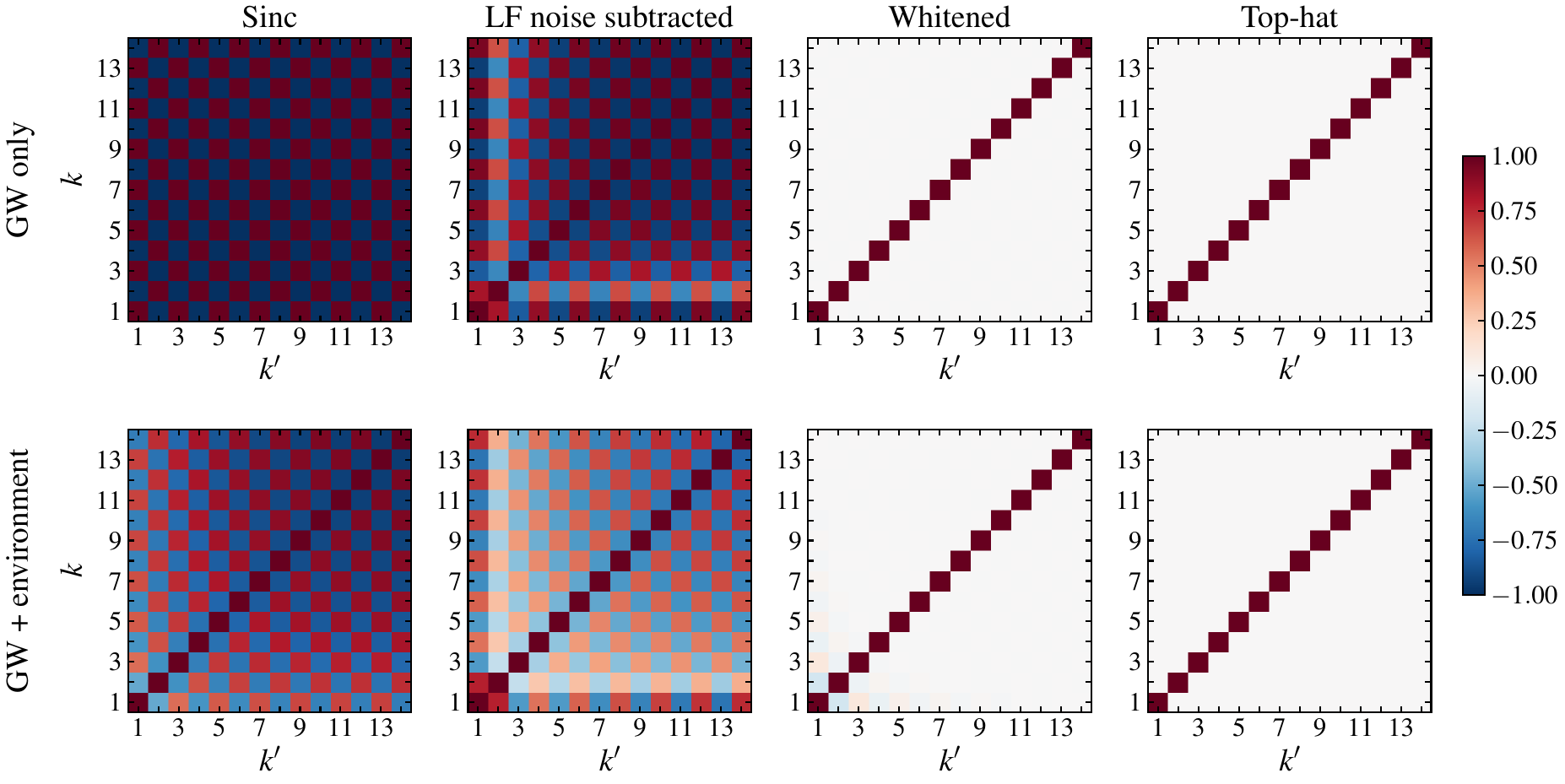}
    \caption{Correlations between the Fourier modes, $\langle \tilde{\delta t}_{J,k} \tilde{\delta t}^{*}_{J,k'} \rangle/\langle |\tilde{\delta t}_{J,k}|^2 \rangle$, calculated from~\eqref{eq:corr_dt}. Columns show results for different window functions, while rows compare cases with and without strong environmental effects.}
    \label{fig:correlations}
\end{figure*}

\subsection{Impact on timing residuals and correlations}

To illustrate the impact of different window functions on inter-mode leakage, the timing residual PDFs and mode correlations are plotted in Figs.~\ref{fig:violins_windows} and~\ref{fig:correlations}, respectively. The correlations, following Eq.~\eqref{eq:corr_dt}, depend exclusively on the window function, which modulates signal leakage across frequencies, and are shown for binaries with and without strong environmental interactions.

The sinc window leads to signal domination by very low-frequency sources, resulting in maximal inter-mode correlations and an RMS spectrum that scales slower than $f^{-1}$. Environmental effects partially mitigate this, but near-maximal correlations persist at higher frequencies. Note also that our $|R|$ is computed in the $fL\gg0.5$ limit, so extremely low-frequency contributions are not properly damped by the antenna response. Therefore, the results of the sinc window using \texttt{GWADpy} are only illustrative. The window function that corresponds to the subtraction of low-frequency noise, Eq.~\eqref{eq:w_substr}, suppresses out-of-band contributions but leaves in-band inter-mode correlations nearly maximal at higher modes, with correspondingly higher timing residuals. Pre-whitening and post-coloring with $\tilde{W}(f) \propto f^{13/6}$~\eqref{eq:w_white} efficiently remove the leakage, yielding PDFs that closely match the top-hat results. In the GW-only scenario, this agreement is exact, since the whitening kernel matches the RMS scaling of that scenario. Small differences emerge in the presence of environmental effects, highlighting the limitations of the linear whitening procedure. The top-hat window thus provides a reasonable approximation of this idealized limit, though some residual leakage is inevitable in practice.

\bibliography{refs}

\end{document}